\documentclass[journal,10pt,draftclsnofoot,onecolumn]{IEEEtran}
\usepackage[english]{babel}
\usepackage[utf8]{inputenc}
\usepackage{listings}
\usepackage[english]{babel}
\usepackage[pdftex]{graphicx}
\usepackage{booktabs}
\usepackage{subfigure}
\usepackage{amsmath}
\usepackage{amsfonts}
\usepackage{amssymb}
\usepackage{mathrsfs}
\usepackage{color}
\usepackage{cite}

{}
{}
{}
\makeatletter
\newif\if@restonecol
\makeatother

\usepackage[linesnumbered,ruled]{algorithm2e}
\usepackage{algpseudocode}
\usepackage{amsmath}

\ifCLASSINFOpdf
\else
\fi

\begin{document}
%
\title{
\huge
Reconfigurable Intelligent Surface-Assisted MAC for Wireless Networks: Protocol Design, Analysis, and Optimization}
%
%
%

\author{Xuelin~Cao,
        Bo~Yang,~\IEEEmembership{Member,~IEEE,}
        Hongliang~Zhang,~\IEEEmembership{Member,~IEEE,}
        Chongwen Huang,~\IEEEmembership{Member,~IEEE,}
        Chau~Yuen,~\IEEEmembership{Fellow,~IEEE,}
        and~Zhu~Han,~\IEEEmembership{Fellow,~IEEE}
        
\thanks{X. Cao, B. Yang, and C. Yuen are with Engineering Product Development Pillar, Singapore University of Technology and Design, Singapore. (E-mail: xuelin$\_$cao, bo$\_$yang, yuenchau@sutd.edu.sg)}
\thanks{H. Zhang is with Department of Electrical Engineering, Princeton University, NJ, USA. (Email: hongliang.zhang92@gmail.com).}
\thanks{C. Huang is with College of Information Science and Electronic Engineering, Zhejiang University, Hangzhou 310027, China, and with International Joint Innovation Center, Zhejiang University, Haining 314400, China, and also with Zhejiang Provincial Key Laboratory of Info. Proc., Commun. $\&$ Netw. (IPCAN), Hangzhou 310027, China. (E-mail: chongwenhuang@zju.edu.cn)}
\thanks{Z. Han is with the Department of Electrical and Computer Engineering in the University of Houston, Houston, TX 77004 USA, and also with the Department of Computer Science and Engineering, Kyung Hee University, Seoul, South Korea, 446-701. (Email: hanzhu22@gmail.com).}
}

\maketitle

\begin{abstract}
Reconfigurable intelligent surface (RIS) is a promising reflective radio technology for improving the coverage and rate of future wireless systems by reconfiguring the wireless propagation environment. The current work mainly focuses on the physical layer design of RIS. However, enabling multiple devices to communicate with the assistance of RIS is a crucial challenging problem. Motivated by this, we explore RIS-assisted communications at the medium access control (MAC) layer and propose an RIS-assisted MAC framework. In particular, RIS-assisted transmissions are implemented by pre-negotiation and a multi-dimension reservation (MDR) scheme. Based on this, we investigate RIS-assisted single-channel multi-user (SCMU) communications. Wherein the RIS regarded as a whole unity can be reserved by one user to support the multiple data transmissions, thus achieving high efficient RIS-assisted connections at the user. Moreover, under frequency-selective channels, implementing the MDR scheme on the RIS group division, RIS-assisted multi-channel multi-user (MCMU) communications are further explored to improve the service efficiency of the RIS and decrease the computation complexity. Besides, a Markov chain is built based on the proposed RIS-assisted MAC framework to analyze the system performance of SCMU/MCMU. Then the optimization problem is formulated to maximize the overall system capacity of SCMU/MCMU with energy-efficient constraint. The performance evaluations demonstrate the feasibility and effectiveness of each.
\end{abstract}


%
\IEEEpeerreviewmaketitle

\section{Introduction}

%
%
%
%
\IEEEPARstart{T}he emergence of the Internet-of-Everything (IoE) system, the rapid development of data-centric and automated processes have extended the coverage of mobile communications from the legacy inter-personal communication to smart inter-connection among millions of people and billions of devices. These devices are spurring worldwide activities focused on defining the next-generation wireless communications that can truly meet the demands for a fully connected and intelligent digital world \cite{letaief2019roadmap,Saad2019vision}. Currently, important 5G and beyond technologies such as Internet-of-Thing (IoT) paradigm have received much attention in academia and industry, which brings considerable challenges and opportunities in the fundamental architecture, the performance of the component, and especially for the standards at the medium access control (MAC) layer \cite{aijaz2020private,aijaz2020high}. It is admitted that the ongoing unprecedented proliferation of intelligent communication is posing new challenges to MAC that needs to cater to low energy consumption, low cost, high intelligence, high capacity, and easy implementation. 

Recently, the reconfigurable intelligent surface (RIS) that also is known as intelligence reflecting surface or holographic multiple input multiple output surface (HMIMOS)\cite{huang2020holographic,huang2020reconfigurable} has been envisioned as intrinsic components of beyond 5G wireless systems to offer unprecedented spectral efficiency gains and drive the wireless architectural evolution \cite{ntontin2019reconfigurable,MA,di2020smart}. The core concept of RIS is stacking a massive number of passive radiating elements to realize a continuous electromagnetically active surface. RIS technology has been verified as an alternative to beamforming techniques since RIS can tune the phase of the incident signal by the intelligent reflection of elements without buffering or processing \cite{hum2013reconfigurable}. In that way, the reflected signal from RIS can be aligned with the non-reflected signal at the receiver to enhance the communication performance and improve the coverage of the wireless systems \cite{nadeem2019large,di2019smart,di2020hybrid,mehrotra20193d,
li2019towards,hu2018beyond}. In other words, manipulating the reflection characteristics of the RIS can improve the propagation condition, enrich the communication channel, and increase the energy efficiency.  

The most existing works focused on the RIS beamforming design considering the global channel state information (CSI)\cite{basar2019large, basar2019transmission, zhang2020reconfigurable, di2020practical, wu2018intelligent, guo2020weighted, huang2019reconfigurable, huang2018achievable, huang2018energy, abeywickrama2020intelligent, han2019large}. In \cite{basar2019large,basar2019transmission}, the authors proposed the RIS-aided transmission strategy index modulation to improve the system spectral efficiency. The authors in \cite{wu2018intelligent} studied the joint optimization problem of active and passive beamforming. The authors in \cite{guo2020weighted} studied an RIS-aided multi-user system to maximize the weighted sum-rate by joint designing the beamforming and the RIS phase-shift. In \cite{huang2019reconfigurable, huang2018achievable, huang2018energy}, the authors aimed to maximize the sum-rate and energy efficiency of RIS-assisted multi-user downlink system. The authors in \cite{abeywickrama2020intelligent} investigated the RIS system that can capture the phase-dependent amplitude variation in the element-wise reflection coefficient. The authors in \cite{han2019large} evaluated the performance of a LIS-assisted large-scale antenna system with different propagation scenarios. Even more recently, RIS-assisted uplink multi-user system and RIS-assisted physical-layer security schemes were discussed in \cite{taha2019enabling,yang2021Int,yu2020robust}. Besides the theoretic analysis of RIS, the RIS prototype has been developed in \cite{LDai}, which combines the functions of phase shift and radiation together on an electromagnetic surface, positive intrinsic-negative (PIN) diodes are used to realize 2-bit phase shifting for beamforming.

\subsection{Challenges and Motivations}
Different from the various existing MAC \cite{aijaz2013prma,shahin2018hybrid,liu2014design}, we focus on designing MAC with the assistance of the RIS. It is generally known that the achievable RIS techniques in the physical layer have shown a profound potential in future wireless communications, which inspires a feasibility solution on RIS enabling in the MAC layer. However, utilizing RIS to serve multiple users’ transmissions at the MAC layer still faces some challenges, such as 1) how to construct RIS-assisted communications for the multi-user system; 2) how to avoid interference at the receiver caused by RIS reflection; 3) how to jointly optimize the RIS configuration and MAC layer parameters. These issues cannot be addressed by directly applying the existing MAC schemes, to the authors’ best knowledge.

Motivated by the challenges raised, in this paper, we mainly focus on the compatible design of RIS-assisted MAC by taking the bellow measures.

\begin{itemize}
\item To address challenge 1), we design a novel RIS-assisted MAC framework, where the negotiation and data transmission are fit together in a hybrid way. During the negotiation phase, each user acquires the RIS resources to assist its data transmissions in a distributed way. This kind of design is especially applicable in the IoT system that pursues spectrum-efficient and energy-efficient.
\item To address challenge 2), we propose the multi-dimension reservation (MDR) scheme for each user. Using the MDR scheme, each user can reserve the RIS resources in advance to achieve multiple RIS-assisted data transmissions without interference during the transmission phase, thus improving the RIS elements utilization.
\item To address challenge 3), we formulate a joint optimization problem for RIS-assisted communications with considering RIS configuration, transmission power, and MAC parameters to maximize the system capacity. 
\end{itemize}

Moreover, the proposed MAC framework is feasibly implemented in practice. Specifically, it is known that the mature medium access schemes (e.g., carrier-sense multiple access with collision avoidance (CSMA/CA) and time-division multiple access (TDMA)) have already been implemented in the commercial off-the-shelf (COTS) products, which may not be likely modified. Therefore, the proposed MAC framework can use the RIS elements efficiently with low complexity, which shows the significance of practical implementation.

	\begin{figure}[t]
      \centering
    	\subfigure[RIS-assisted SCMU communications]{
    		\begin{minipage}[b]{0.39\textwidth}
    			\includegraphics[width=1\textwidth]{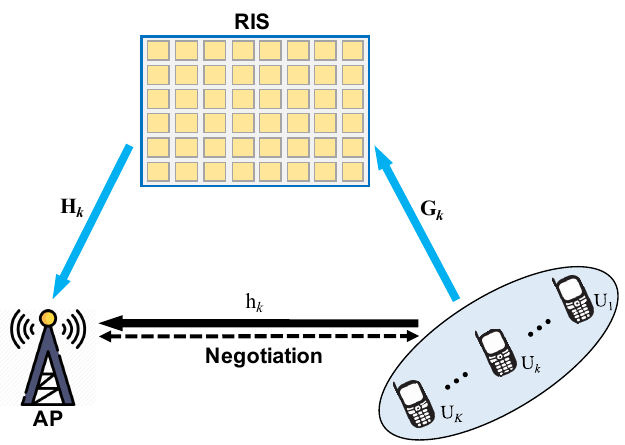}
    		\end{minipage}
    	}
    	\subfigure[RIS-assisted MCMU communications]{
    		\begin{minipage}[b]{0.39\textwidth}
    			\includegraphics[width=1\textwidth]{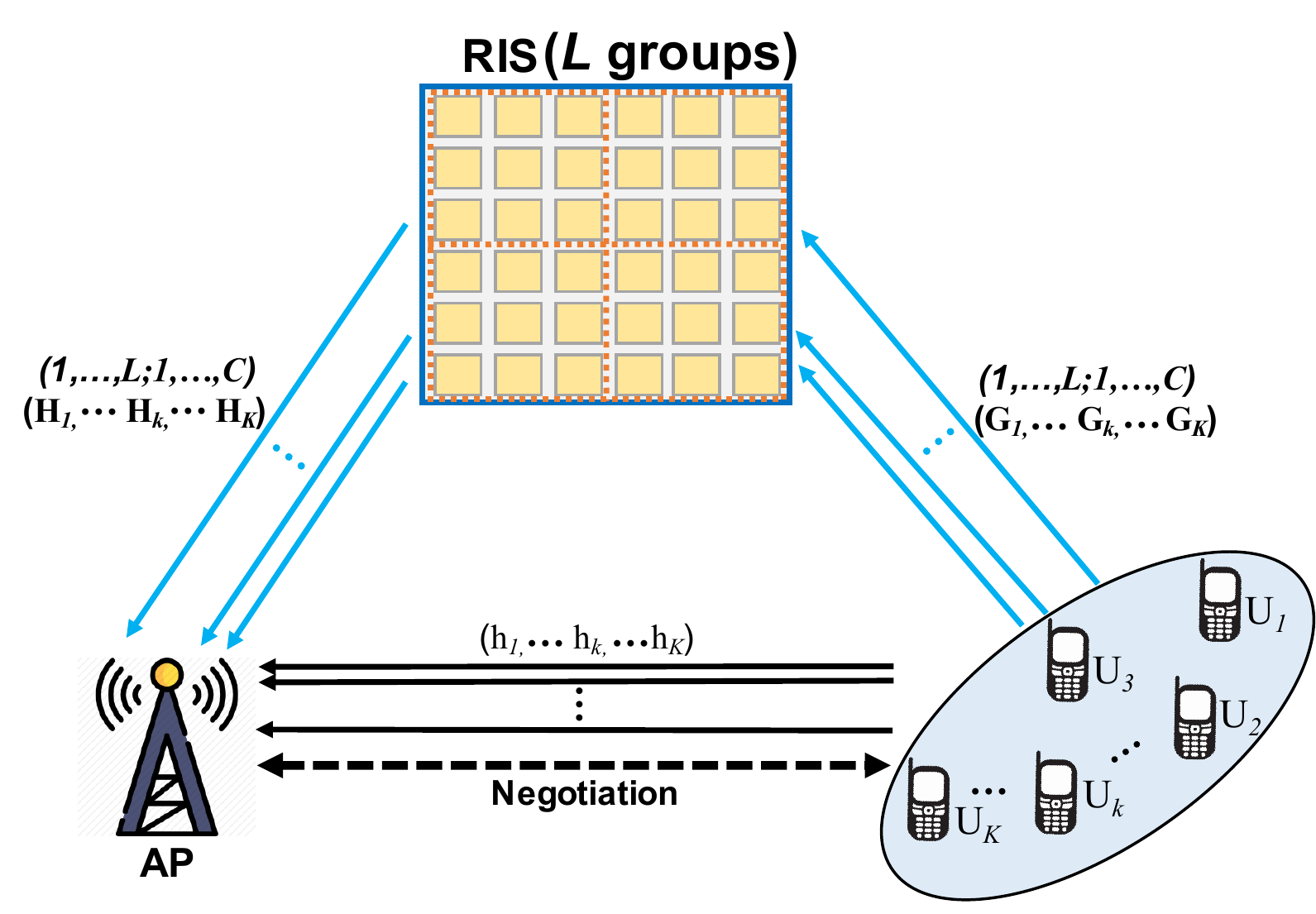}
    		\end{minipage}
    	}
    	\caption{RIS-assisted SCMU/MCMU communications system model.} 
    	\label{sys}
    \end{figure} 

\subsection{Contributions}
In the paper, the main contributions are summarized as follows.

\begin{itemize}
	\item \textit{An RIS-assisted MAC framework.} We propose an RIS-assisted MAC framework for the multi-user uplink system. In this framework, we separate the negotiation and RIS-assisted data transmissions in the time domain using the MDR scheme, and we solve the RIS elements allocation issue at the MAC layer to avoid interference.
	
	\item \textit{RIS-assisted SCMU/MCMU communications.} We explore RIS-assisted single-channel multi-user (SCMU) communications based on the proposed RIS-assisted MAC framework. With the implementation of the SCMU, the RIS serves one user at a time for completing the energy-efficient transmission. By introducing the RIS group division, we then further extend to RIS-assisted multi-channel multi-user (MCMU) communications to support concurrent transmissions via the RIS.	
	
    \item \textit{The analysis of RIS-assisted SCMU/MCMU communications.} We construct a Markov model to analyze the system performance of RIS-assisted SCMU/MCMU communications. Unlike the existing analysis, we separate the contention process and RIS-assisted data transmissions in the time domain. Accordingly, we introduce the contention state and the transmission state, which correspond to the negotiation period and the data transmission period of the proposed RIS-assisted MAC framework, respectively.
    
    \item \textit{The optimization and simulation results.} We formulate the joint optimization problems to maximize the system capacity of RIS-assisted SCMU/MCMU communications with considering RIS configuration, transmit power, and MAC parameters. Moreover, we conduct the simulations on RIS-assisted SCMU/MCMU communications and demonstrate the performance of both in terms of system throughput, signal-to-noise ratio (SNR), and transmit power.
\end{itemize}

\subsection{Organization and Notations}
The rest of this paper is organized as follows. In Section \ref{sec2}, RIS-assisted communications system model is presented. Section \ref{sec2c} proposes an RIS-assisted MAC framework. Section \ref{sec3} and \ref{sec4} investigate how to implement the proposed RIS-assisted MAC framework in SCMU communications and MCMU communications, respectively. In Section \ref{sec6}, performance evaluation is presented. Finally, Section \ref{sec7} concludes the paper. \textbf{Notation}: $\mathbb{C}^{x*y}$ represents a $x*y$ complex-valued matrix. $\text{diag}(\mathbf{z})$ is a diagonal matrix with its diagonal elements given in vector $\mathbf{z}$.

 
\section{RIS-assisted Communications System Model}\label{sec2}
In this section, the system scenario, the channel model, and the propagation model are presented, where the PHY layer aspects are introduced since we design the suitable MAC based on the RIS physical layer characteristics \cite{basar2019wireless,yildirim2019propagation}.

\subsection{System Scenario}\label{sec2a}
We consider a multi-user uplink wireless system, where the RIS is attached with a RIS controller that can be controlled by the AP via a separate wireless link \cite{huang2019reconfigurable}. The RIS consists of $N$ passive elements and can be further divided into $L$ groups. The $L$ RIS groups are bonded with $C$ sub-channels that can be used to assist the communication of $K$ users (i.e., U$_k, k\in[1, K]$) with one AP. The channels from the U$_k$ to the RIS, from the RIS to the AP, and from the U$_k$ to the AP are denoted by the set $\mathbf{G}_k=\{G_k^1,\ldots,G_k^n,\ldots,G_K^{N/L}\}$, the set $\mathbf{H}_k=\{H_k^1,\ldots,H_k^n,\ldots,H_K^{N/L}\}$, and $h_k$, respectively, where $\mathbf{G}_k\in \mathbb{C}^{(N/L)\times1}$ and $\mathbf{H}_k\in \mathbb{C}^{1\times(N/L)}$. Besides, the communication bandwidth is $B$, and the occupied bandwidth by each sub-channel is $B/C$. We mainly focus on the uplink channel access of multiple users with the assistance of RIS. Note that each RIS element is just allowed to support one user's transmission and reflect its signal to AP at one time. Otherwise, interference occurs. According to the value of $C$ and $L$, the system model in Fig. \ref{sys} can be classified into two cases as follows: when $C=1$ and $L=1$, RIS-assisted SCMU communications are shown in Fig. \ref{sys} (a), where all RIS elements serve one user at a time on one channel. On the other hand, as $C>1$ and $L>1$, RIS-assisted MCMU communications are shown in  Fig. \ref{sys} (b), where $L$ RIS groups serve multiple users at a time on $C$ sub-channels.

\subsection{Channel Model}\label{sec2a1}
Based on the RIS channel model, the received signal at AP from U$_k$ in RIS-assisted SCMU/MCMU communications can be denoted by
\begin{equation}\label{received signal}
\centering
y_k^{SCMU}=\underbrace{h_ks_k}_{\text{Direct data link}}+\underbrace{{\bf H}_k {\bf\Theta}_k  {\bf G}_k s_k}_{\text{RIS-assisted data link}}+w_k 
\end{equation} 
and
\begin{equation}\label{received signal1}
\centering
y_k^{MCMU}=\underbrace{h_ks_k}_{\text{Direct data link}}+\underbrace{{\bf H}_k^l {\bf\Theta}_k^l  {\bf G}_k^l s_k}_{\text{RIS-groups-assisted data link}}+w_k, 
\end{equation}
where $s_k$ represents the signal from U$_k$, which is identically distributed (i.i.d.) random variable with zero mean and unit variance, $w_k$ is additive white Gaussian noise (AWGN) of U$_k$ at AP with zero mean and variance $\sigma^2$. For RIS-assisted transmission channel in \eqref{received signal} and \eqref{received signal1}, ${\bf\Theta}_k$ and ${\bf\Theta}_k^l$ can be defined as
\begin{equation}\label{channel}
\centering
{\bf\Theta}_k=\text{diag}(\phi_k^1,\ldots,\phi_k^n,\ldots,\phi_k^N)
\end{equation} 
and
\begin{equation}\label{channel1}
\centering
{\bf\Theta}_k^l=\text{diag}(\phi_k^{l(1)},\ldots,\phi_k^{l(n)},\ldots,\phi_k^{l(N/L)}).
\end{equation}

Let $\phi_k^n=\beta_k^ne^{j\theta_k^n}, n\in[1,N]$ and $\phi_k^{l(n)}=\beta_k^{l(n)}e^{j\theta_k^{l(n)}}, {l(n)}\in[1,N/L]$ denote the phase-shift of each element $n$ on the RIS group $l$ for U$_k$, where $\{\theta_k^n,\beta_k^n\}$ and $\{\theta_k^{l(n)},\beta_k^{l(n)}\}$ are the phase shift and amplitude reflection coefficient of element $n$ on the RIS group $l$ for U$_k$, respectively. In practice, considering the hardware implementation, we assume that the discrete phase shift with constant amplitude is operated on each RIS element. Then, we have 
\begin{equation}
\phi_k=\lbrace\phi_k^n\vert\phi_k^n=e^{j\theta_k^n}, \theta_k^n \in \mathbf{\Omega},\forall n\rbrace,
\end{equation}
and 
\begin{equation}
\phi_k^{l(n)}=\lbrace\phi_k^{l(n)}\vert\phi_k^{l(n)}=e^{j\theta_k^{l(n)}}, \theta_k^{l(n)} \in \mathbf{\Omega},\forall n\rbrace.
\end{equation}
Without loss of generality, denote by $b$ the number of bits used to represent each of the levels, and then the set of phase-shifts at each element, $\mathbf{\Omega}$, is given by $\mathbf{\Omega} = \{0, \Delta\theta, \ldots, \Delta\theta(\Psi-1)\}$, where $\Delta\theta = 2\pi/\Psi$, and $\Psi = 2^b$ denotes the number of quantized reflection coefficients values of RIS elements. To achieve the best signal reflection, the RIS controller should adjust its reflection coefficient to make the reflected signals coherently added at AP deliberately. 

Let ${\rho_k}^2$ be the transmitter power from U$_k$ to AP, the received SNR at the AP from U$_k$ in RIS-assisted SCMU/MCMU communications can be calculated as 
\begin{equation}\label{SNRE}
\centering
\text{SNR}_k^{SCMU}=\left|\left(h_k+{\bf H}_k {\bf\Theta}_k {\bf G}_k \right)\rho_k\right|^2/\sigma^2 
\end{equation} 
and
\begin{equation}\label{SNRE1}
\centering
\text{SNR}_k^{MCMU}=\left|\left(h_k+{\bf H}_k^l {\bf\Theta}_k^l {\bf G}_k^l\right)\rho_k\right|^2/\sigma^2. 
\end{equation} 

\begin{figure}
	\centering{\includegraphics[width=0.5\textwidth]{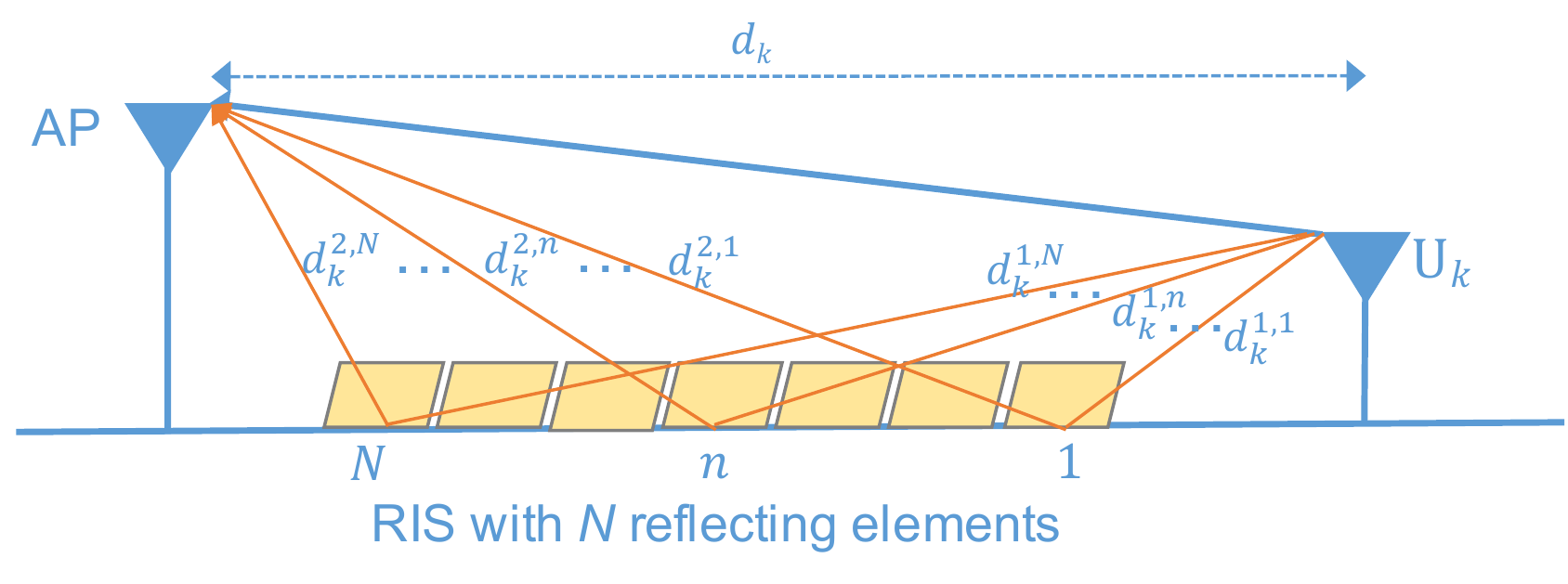}}
	\caption{The reflection-based propagation model via the RIS.}
	\label{mod}
\end{figure}

\subsection{Propagation Model}\label{sec2b}
The motivation of the RIS is to control the radio environment, where the wireless channel can be turned into a deterministic by reconfiguring the propagation of the electromagnetic wave in a software-controlled fashion. The reflection-based propagation model via the RIS in a free-space environment is shown in Fig. \ref{mod}. The received power at AP from U$_k$ through the RIS link can be obtained as
\begin{equation}\label{RP}
\Gamma_k = {\rho_k}^2 \left(\frac{\lambda}{4 \pi}\right)^2 \left|\frac{1}{d_k}+\sum_{n=1}^{N}\frac{\phi_k^n\times e^{-j\Delta\theta_k^n}}{d_k^{1,n}+d_k^{2,n}}\right|^2,
\end{equation}
where $\lambda$ is the wavelength, $d_k$ is the distance of direct link from U$_k$ to AP. The distance between the U$_k$ and the RIS element $n$, and the RIS element $n$ and AP are denoted by $d_k^{1,n}$ and $d_k^{2,n}$, respectively. Moreover, $\Delta{\theta_k^n}=\frac{2\pi\left(d_k^{1,n}+d_k^{2,n}-d_k\right)}{\lambda}$ is the phase difference between the RIS link and the direct link. When $d_k$ is sufficiently large, we can assume $d_k=d_k^{1,n}+d_k^{2,n}, \forall n\in[1,N]$. To achieve the best performance of the RIS, the phase-shift of each RIS element is aligned with the phase of the LOS path. Considering this ideal case, \eqref{RP} can be re-written as below
\begin{equation}
\Gamma_k \approx \left(N+1\right)^2 {\rho_k}^2 \left(\frac{\lambda}{4 \pi d_k}\right)^2,
\end{equation}
then the received power at AP from U$_k$ can be simplified as follows
\begin{equation}
\Gamma_k \propto N^2 {\rho_k}^2 \left(\frac{1}{d_k}\right)^2,
\end{equation}
which shows that the received power at the AP is proportional to $N^2$ and is inversely proportional to $d_k^2$. In other words, the number of RIS elements can affect the power gain and thus displays potential in low power and data-intensive wireless communications. Note that $\Gamma_k \propto {\rho_k}^2 \left(\frac{1}{d_k}\right)^2$ holds if RIS is not used, i.e., the part that $\sum_{n=1}^{N}\frac{\phi_k^n\times e^{-j\Delta\theta_k^n}}{d_k^{1,n}+d_k^{2,n}}$ in \eqref{RP} can be removed. This can be viewed as a direct link channel model.

Therefore, the total power consumption of the U$_k$ $\rightarrow$ RIS $\rightarrow$ AP link can be given as 
\begin{equation}
P_{k,c}={\rho_k}^2+P_{RIS}+\Gamma_{k},
\end{equation}
where $P_{RIS}$ is the hardware static power consumption at the RIS, ${\rho_k}^2$ is the transmit power at U$_k$, and $\Gamma_{k}$ is the revived power at AP from U$_k$. If the RIS is not used, only direct link U$_k$ $\rightarrow$ AP exists, the total power consumption of the direct link is $P+\Gamma_{k}$, where $P$ represents the transmit power of any user without using the RIS. In other words, the transmit power $ P $ can enable the AP to receive the signal from any user.

\begin{figure}[t]	
\centering{\includegraphics[width=0.5\textwidth]{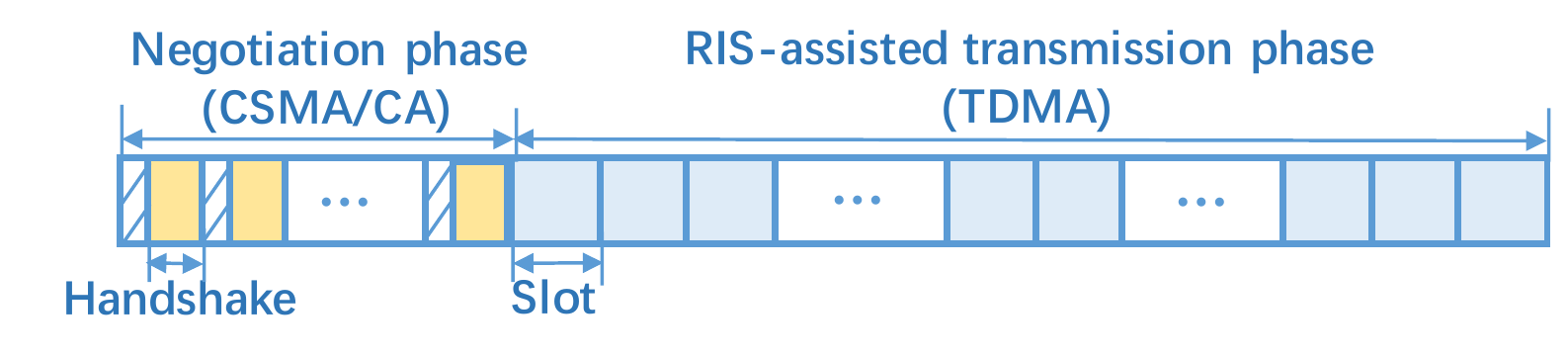}}
	\caption{An RIS-assisted MAC framework.}
	\label{Frame}
\end{figure}

\section{RIS-assisted MAC framework}\label{sec2c}
In the multi-user uplink communications system, an RIS-assisted MAC framework is designed in Fig. \ref{Frame}. Following the CSMA scheme, each user proposes the uplink transmissions and reserves the RIS resources (i.e., the RIS elements). The AP estimates the channels of the direct link and RIS links, and controls the RIS to assist each user's uplink data transmissions in TDMA mode. The detailed work on the channel estimation can refer to the related literature \cite{you2020channel,wei2020parallel}. 
\subsubsection{Frame Structure}
The frame is shown in Fig. \ref{Frame}, where each frame is separated into two phases: the negotiation phase and the RIS-assisted transmission phase \cite{YB}. The RIS-assisted transmission phase can be further divided into a series of slots. In the negotiation phase, each user competes for the access privilege according to the backoff mechanism and then negotiates with the AP to reserve the RIS resources, power resources, channel resources, and slot resources. As the RIS-assisted transmission phase arrives, the RIS can be controlled and switched to the requested users on the negotiated channels and slots.

\subsubsection{MDR Scheme}
Multi-dimension reservation is proposed in the negotiation phase to implement RIS-assisted data transmissions. Each dimension is defined as one type of reservation information.
\begin{itemize}
\item Dimension 1 (D1): time zone. The slots that are in the RIS-assisted transmission phase can be reserved on the channel;
\item Dimension 2 (D2): power zone. The transmit power can be used at the requested user while starting RIS-assisted data transmissions; 
\item Dimension 3 (D3): RIS zone. The phase-shift of each RIS element, which can be controlled by the AP to align with the user's direct data transmission in the RIS-assisted transmission phase.
\item Dimension 4 (D4): frequency zone. The sub-channel and the corresponding RIS group are requested by the user at the negotiation phase for data transmission at the RIS-assisted transmission phase. Note that one sub-channel is bonded with one RIS group. Once the sub-channel is reserved, the accordingly RIS group is reserved;
\end{itemize}

\subsubsection{Extended Control Packets}
Two types of extended control packets are introduced to implement the MDR scheme, i.e., extended request-to-send (eRTS) and extended clear-to-send (eCTS). The control packet structure of each is shown as follows:
\begin{itemize}
\item eRTS (${T_k}^{ini}, r_k, T_{cycle}, c_k, l_k$) of U$_k, k\in[1, K]$. eRTS is an extension of the traditional RTS packet defined in IEEE 802.11 DCF. Compared to RTS, five additional fields, i.e., the initial transmission time (${T_k}^{ini}$), the transmission cycle ($T_{cycle}$), the number of reserved transmissions ($r_k$), the sub-channel ($c_k$) and the RIS group ($l_k$), are added into the eRTS. Note that the fields of $c_k$ and $l_k$ are omitted in RIS-assisted SCMU communications.
\item eCTS (${\rho_k}^2$, ${T_k}^{ini}$, $r_k$, $T_{cycle}, c_k, l_k$) of U$_k, k\in[1, K]$. eCTS is an extension of the traditional CTS packet defined in IEEE 802.11 DCF. Compared to CTS, six additional fields, i.e., the transmission power(${\rho_k}^2$), ${T_k}^{ini}$, $T_{cycle}$, $r_k$, $c_k$ and $l_k$, are added into the eCTS. Note that the fields of $c_k$ and $l_k$ are omitted in RIS-assisted SCMU communications.
\end{itemize}

\begin{figure*}[t]	
    \centering{\includegraphics[width=1\textwidth]{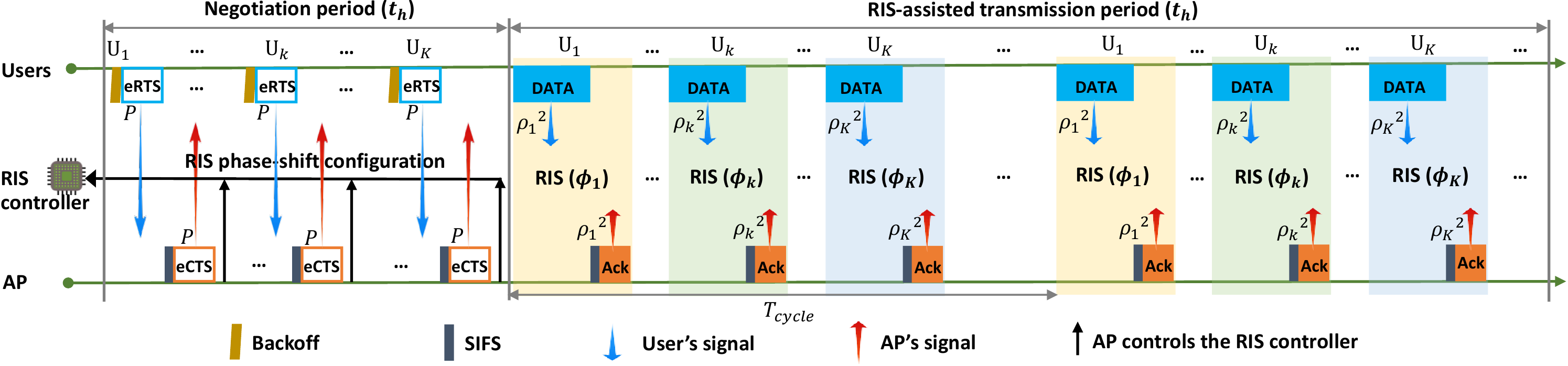}}
	\caption{RIS-assisted SCMU communications, where $L=1$ and $C=1$. Each user contends the access privilege and negotiates with AP (i.e., eRTS/eCTS exchanging) in a single channel.}
	\label{pro}
\end{figure*}

\section{RIS-assisted SCMU Communications}\label{sec3}
In this section, as shown in Fig. \ref{sys} (a), we explore how to implement RIS-assisted SCMU communications for the multi-user uplink communications system, where only one channel is considered, and all RIS elements serve one user (i.e., $L=1$ and $C=1$).
\subsection{Basic Idea}\label{sec3a}

The proposed RIS-assisted SCMU communications are shown in Fig. \ref{pro}. Based on the proposed RIS-assisted MAC framework, the three-dimension reservation that includes D1, D2, and D3 is introduced in the negotiation phase, wherein each user competes for the channel to reserve the available slots and RIS elements with AP. AP calculates the RIS reflection coefficients and the transmit power for the requested user and then informs them of the data transmission during the RIS-assisted transmission period. Once the mentioned three-dimension reservation is constructed, the requested user commences its transmissions in the reserved slots with the reserved transmit power, and the RIS controller adjusts the phase-shift of each RIS element at the reserved slots to assist the requested user's transmissions. In RIS-assisted SCMU communications, since the RIS controller updates all RIS elements' reflection coefficients for one user at one time, the complexity of RIS phase-shift adjustment is $\mathcal{O}(N)$.

\subsection{Implementation Algorithm}\label{sec3c}
The implementation algorithm of RIS-assisted SCMU communications is presented in Algorithm \ref{algo}. Considering the uplink traffics of U$_k, k\in[1, K]$, Algorithm \ref{algo} contains two phases:
\subsubsection{Phase 1: Negotiation Phase}
\begin{itemize}
     \item Step 1: The requested user competes for the channel by operating the backoff scheme during the negotiation period, i.e., each user sets its backoff value $\{W, s\}$, where $W\in[CW_{min}, CW_{max}]$ and $s\in[0,m]$ denote the backoff counter and the backoff stage, respectively. $CW_{min}$ and $CW_{max}$ are the minimum contention window and maximum contention window, $m$ is the maximum backoff stage. Once the channel is sensed to be idle, each user decreases $W$ by one after a slot. Once the user decreases $W$ to zero, it wins the channel access privilege. If a collision occurs, the value of $s$ increases by one until $s=m$.   
	\item Step 2: By sending the eRTS packet to the AP with the power $P$, U$_k$ requests its $T_k^{ini}$, $T_{cycle}$, and $r_k$ from AP for its multiple RIS-assisted transmissions. Besides, the other users keep silent by setting the network allocation vector (NAV) when U$_k$ is negotiating with the AP. 
	\item Step 3: On receiving the eRTS packet, if no collision occurs on the reserved RIS and slots, the AP estimates the phase-shift of each RIS element, $\phi_k$, and calculates the transmit power, ${\rho_k}^2$, at U$_k$ as well. Besides, the AP records ${\rho_k}^2$, ${T_k}^{ini}$, $r_k$ and $T_{cycle}$ into the eCTS packet, and replies it to U$_k$ while notifying the $\phi_k$, ${T_k}^{ini}$, $r_k$ and $T_{cycle}$ to the RIS controller. Otherwise, the AP uses a new value ${T_k^{ini}}^*$ instead of ${T_k}^{ini}$ in the eCTS packet, and replies it to U$_k$ while notifying the RIS controller. 	
	\item Step 4: Once U$_k$ receives the eCTS packet from the AP, U$_k$ ceases its competition in the current negotiation period and waits for the RIS-assisted transmission phase. When the NAV expired out, other requested users continue the backoff procedure until the negotiation period terminates.   
	\end{itemize}
\subsubsection{Phase 2: RIS-assisted Transmission Phase}
\begin{itemize}
     \item Step 1: When the reserved slots arrive, U$_k$ sends the data with the transmit power, ${\rho_k}^2$. Meanwhile, RIS controller sets the phase-shift of each RIS element as $\theta_k^n$, $\theta_k^n\in\mathbf{\Omega},\forall n\in[1,N]$, and reflects U$_k$'s signal to the AP.
	\item Step 2: On receiving the accumulated reflected signal from the RIS and the LOS's signal from U$_k$ correctly, the AP replies an ACK packet to U$_k$.  
	\item Step 3: During RIS-assisted transmissions between U$_k$ and AP, other users set the NAV to cease their transmissions until the reserved RIS-assisted transmissions of U$_k$ finishes.
\end{itemize}

\begin{algorithm}
    \caption{RIS-assisted SCMU communications}
    \label{algo}
    \small
    Initialize $\vert\phi_k^n\vert = 0$, ${\rho_k}^2=P$, $\forall j, k\in[1,K]$\;
    $t = 0$\;
    \textbf{Phase 1: Negotiation phase}\;
    \If{$t$ is in \textbf{Phase 1}}
    {
        \If{U$_k$ has data to transmit}
        {U$_k$ sets its backoff value $\{W, s\}$\;
            \If{$W=0$ }
             {U$_k$ sends eRTS to AP\;
             U$_{j, j\neq k}$ sets NAV and keeps silent\;
             AP receives eRTS and judges\; 
                 \If{no collision} 
                 {AP calculates $\phi_k, {\rho_k}^2$\; 
                  AP replies eCTS to U$_k$\;
                  Ap notifies the RIS controller about the $\phi_k,{T_k}^{ini},r_k,T_{cycle}$\;
                 }
                 \Else
                 {AP replies eCTS to U$_k$\;
                  AP notifies the RIS controller about the $\phi_k,{{T_k}^{ini}}^*,r_k,T_{cycle}$\;
                 }
                U$_k$ receives eCTS and waits for \textbf{Phase 2}\;  
             }
              \Else
              {$W--$\; 
              }
      }
         \Else
              {U$_{k}$ keeps silent\;
              }
    }  
         \Else
              {U$_{k}$ waits for \textbf{Phase 1}\;
              }   
     \textbf{Phase 2: RIS-assisted transmission phase}\;   
      \If{$t$ is in \textbf{Phase 2}}
      {U$_{k}$ transmits data to AP at the reserved slots with power ${\rho_k}^2$\;
       RIS controller sets $\phi_k$ and reflects U$_k$'s signal\;
       AP receives reflected and directed signals from U$_k$\;
       AP replies ACK to U$_{k}$\; 
      }  
      \Else
              {U$_{k}$ waits for \textbf{Phase 2}\;
              }   
\end{algorithm}

\subsection{Performance Analysis}\label{sec3d}
In this section, a two-state transition model \cite{cao2018performance} is illustrated for the proposed RIS-assisted MAC framework, where each user maintains two states: the contention state ``C” and the transmission state ``T”. Let $\gamma$ and $\eta$ denote the transmission rate from ``C” to ``T” and from ``T” to ``C”, respectively. Then, the stationary probability of two states can be given as
\begin{equation}\label{e1}
\mho \buildrel \Delta \over = \left(q,1-q\right) = \left(\frac{\eta }{\gamma  + \eta},\frac{\gamma }{\gamma  + \eta}\right),
\end{equation}
where $q$ and $1-q$ represent the stationary probability that a user keeps in the state of ``C” and ``T”, respectively.

\begin{figure}[t]
	\centering{\includegraphics[width=0.5\textwidth]{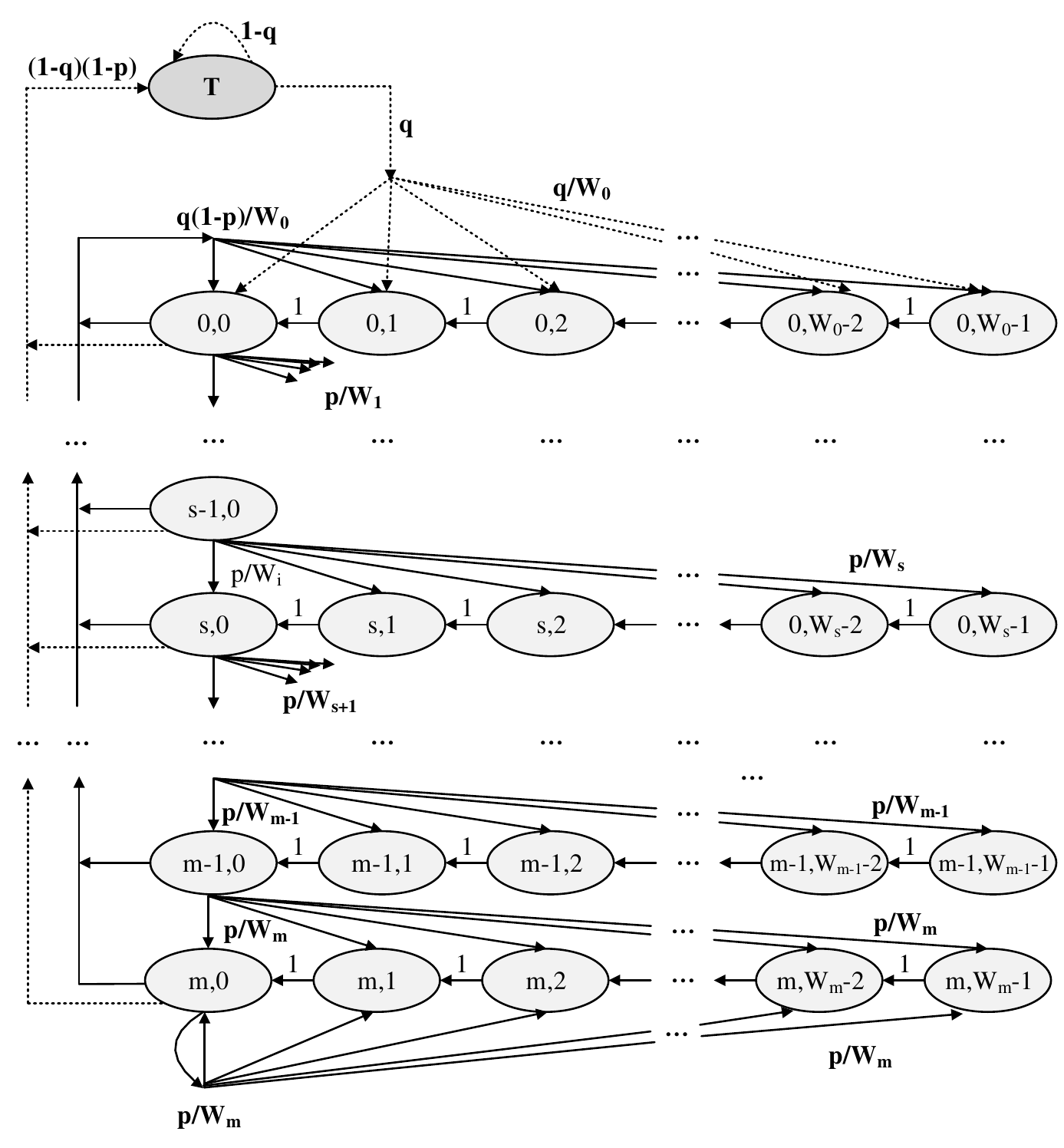}}
	\caption{Enhanced discrete Markov model based on the two-state transition.}
	\label{Mode}
\end{figure}

The two-state transition model, combined with the traditional backoff process, is constructed in Fig. \ref{Mode}, which is different from the Bianchi's model \cite{bianchi2000performance} since the contention and the data transmission are separated. Specifically, the state ``C” corresponds to the negotiation period, and the state ``T” corresponds to the data transmission period of the proposed RIS-assisted MAC framework. In our model, each contention state of user is denoted by $(s,W)$, where $s \in [0,m]$ and $W \in [0,W_m]$ denote the backoff stage and the backoff counter, respectively. Then, we can get
\begin{equation}\label{e2}
\centering
\left\{ \begin{array}{l}
1)\ {P_{s,W|s,W + 1}} = 1,\;\;\;\;\;\;\;\;\;\;\;\; W \in [0,{W_s} - 2],\;s \in [0,m],\\
2)\ {P_{0,W|s,0}} = q(1 - p)/{W_0}, W\in [0,{W_0} - 1],\;s \in [0,m],\\
3)\ {P_{s,W|s - 1,0}} = p/{W_s},\;\;\;\;\;\;\; W \in [0,{W_s} - 1],\;s \in [1,m],\\
4)\ {P_{m,W|m,0}} = p/{W_m},\;\;\;\;\;\;\; W\in [0,{W_m} - 1],\\
5)\ {P_{T|s,0}} = (1 - q)(1 - p),\; s \in [0,m],\\
6)\ {P_{0,W|T}} = q/{W_0},\;\;\;\;\;\;\;\;\;\;\;\; W \in [0,{W_0} - 1],\\
7)\ {P_{T|T}} = 1 - q,\;\;\;
\end{array} \right.
\end{equation}
where $p$ is the probability that a collision happen while one user transmits. $W_s=2^sW_0$ is the contention window when backoff stage is $s$, where $W_0$ denotes the minimum contention window. The expressions $1) \sim  4)$ in \eqref{e2} describe the backoff process when users stay in the state of ``C”, the remaining expressions $5)\sim 7)$ show that the reserved transmission process when users keep the state of ``T”. Specifically, the expression $1)$ means that the backoff counter is decremented one with probability $1$. The expression $2)$ is explained that a new packet following a successful packet transmission begins a new backoff process with probability $q(1-p)/W_0$, here, $s$ is valued $0$, $W$ is reset in the range $\left(0, W_0-1\right)$. The expression $3)$ shows that an unsuccessful transmission occurs at the backoff stage of $s-1$, the backoff stage is increased one (i.e., backoff stage become $s$), and a new initial backoff counter value is uniformly selected in the range $(0, W_s)$. The expression $4)$ illustrates that once $s$ equals to the maximum value $m$, it is not increased in subsequent packet transmissions. The expression $5)$ models the fact that a successful packet transmission switches from contention state to transmission state with probability $(1 - q)(1 - p)$, and the expression $6)$ illustrates the user switches into contention state from transmission state with probability $q/W_0$. Finally, the expression $7)$ shows that users maintain its original state of ``T” with probability $1-q$.
     
In the proposed RIS-assisted MAC framework, since the MDR scheme is introduced into the access process of each user, the stationary probability that one user transmits a data packet in a random slot during the RIS-assisted transmission phase, $\tau$, is given by 
\begin{align}\label{eq7}
	\begin{split}
	\centering
	\tau = \frac{2(1-2p)q}{{q[(\text{CW}_0 + 1)(1 - 2p) + p\text{CW}_0(1 - {{(2p)}^m})] + 2(1 - q)(1 - p)(1 - 2p)}},
	\end{split}
\end{align}
where $p=1-(1-\tau)^{Kq-1}$ is the collision probability and $q = \eta /(\gamma  + \eta)$ is the stationary probability that one user stays in the state of ``C”. 

Here, $\gamma$ and $\eta$ can be calculated as
\begin{equation}\label{rate}
\centering
\left\{\begin{array}{l}
\gamma=\frac{\zeta_s}{\zeta_e\delta+\zeta_st_s +\zeta_ct_c},\\
\eta=\frac{1}{t_pr_{max}},\;\;\;\;
\end{array} \right.
\end{equation}  
where $t_s = eRTS+eCTS+DIFS+SIFS$, $\delta$ and $t_c=eRTS+DIFS$ represent the time duration of successful negotiation, idle state and collision, respectively. Here, $eRTS$ and $eCTS$ are the transmissions time of the eRTS packet and the eCTS packet, respectively. $DIFS$ and $SIFS$ are the duration of DIFS and SIFS, respectively. $t_p$ denotes the time length of one data transmission, and $r_{max}$ is the maximum number of reserved transmission which equals to $\frac{t_r}{T_{cycle}}$. $\zeta_s, \zeta_e$ and $\zeta_c$ represent the probability of successful RIS-assisted transmission, the probability of idle channel and the probability of collision, respectively, which are given as
\begin{equation}\label{Prob}
\centering
\left\{ \begin{array}{l}
\zeta_s = Kq\tau(1-\tau)^{Kq-1},\\
\zeta_e = (1-\tau)^{Kq},\\
\zeta_c = 1-\zeta_s-\zeta_e.\\
\end{array} \right.
\end{equation}

To improve the channel utilization, we suggest $T_{cycle} = N_r * t_{p}$, i.e., a guideline for setting $T_{cycle}$ in the proposed RIS-assisted SCMU/MCMU communications, where $N_r$ can be approximately calculated by $\lfloor\frac{t_h\zeta_s}{t_s}\rfloor$. When $t_p$ is given, a large $T_{cycle}$ means that a big $N_r$, which brings the more overheads. Note that if the number of data packets accumulated at the MAC queue is larger than $r_{max}$, the user needs to contend again in the next negotiation phase. 

According to \eqref{SNRE}, the achievable data rate at U$_k$ can be given by  
\begin{equation}\label{data rate}
\centering
R_k^{SCMU}\left(\rho_k,\phi_k \right)=B\log_2\left( 1+\text{SNR}_k^{SCMU}\right), 
\end{equation} 
where $B$ is the channel bandwidth.

The total capacity of RIS-assisted SCMU communications can be calculated as follow:
\begin{align}\label{data rate1}
\begin{split}
\Im_{Total}^{SCMU}&=\sum_{k=1}^{K}\frac{t_pr_k}{T}\varepsilon_kR_k^{SCMU}\left(\rho_k,\phi_k \right)=\frac{t_p}{T}B\sum_{k=1}^{K}r_k\varepsilon_k\log_2\left( 1+\text{SNR}_k^{SCMU}\right), 
\end{split}
\end{align}
where $r_k$ is the number of data transmissions at U$_k$ in the RIS-assisted transmission phase. $T=t_h+t_r$ are the length of a frame, where $t_h$ and $t_r$ are the length of the negotiation period and the length of the RIS-assisted transmission period, respectively. $t_p$ is the time length of one data transmission, and $t_pr_k$ denotes the time length of $r_k$ data transmissions at U$_k$ in the RIS-assisted transmission phase. $\varepsilon_k=\frac{t_h\zeta_s}{t_sK}$ is the probability that U$_k$ commences its data transmissions during the RIS-assisted transmission period. $\text{SNR}_k^{SCMU}$ is the signal-to-noise ratio at U$_k$ with RIS-assisted SCMU communications. Given the fixed value of $\varepsilon_k$, \eqref{data rate1} can be written as
\begin{align}\label{data rate2}
\begin{split}
\Im_{Total}^{SCMU}&=\frac{t_pt_h\zeta_s}{Tt_sK}B\sum_{k=1}^{K}r_k\log_2\left( 1+\text{SNR}_k^{SCMU}\right). 
\end{split}
\end{align}

\subsection{Performance Optimization}\label{sec3e}
Based on the aforementioned performance analysis of the RIS-assisted MAC framework, the performance optimization for RIS-assisted SCMU communications can be formulated as:
\begin{align}\label{p1}
\begin{split}
\textbf{P1}:\ &\mathop {\max}\limits_{{\rho_1}^2,\ldots,{\rho_K}^2;{\bf\Theta}_1,\ldots,{\bf\Theta}_K;r_1\ldots,r_K}\Im_{Total}^{SCMU}\\ 
\bf{s.t.}\ \
&\text{C1:}\ \ 0\leq\rho_{k}^2 \leq P-P_{RIS}, \ \forall k,\\
&\text{C2:}\ \ \vert\phi_k^n\vert=1, \ \ \ \forall k, n, \\
&\text{C3:}\ \ \theta_k^n\in \mathbf{\Omega}, \ \ \ \ \forall k, n,\\
&\text{C4:}\ \ t_h+t_r = T, \ \ \ \forall k\\
&\text{C5:}\ \ \frac{t_p}{t_s}r_{max}\zeta_s\leq\frac{t_r}{t_h}, \ \ \ \forall k\\
&\text{C6:}\ \ 1 \leq r_{k}\leq r_{max}, \ \ \ \ \forall k,
\end{split}
\end{align}
where $\text{C1}$ constrains the transmit power of each user. $\text{C2}$ and $\text{C3}$ separately illustrates the constraints of the amplitude and phase-shift of RIS elements. $\text{C4}$ constrains the length of a frame, and $\text{C5}$ constrains the RIS-assisted transmission period and the negotiation period. $\text{C6}$ limits the number of data transmissions.

In \eqref{data rate1}, since $\frac{t_pt_h\zeta_s}{Tt_sK}B$ is constant, problem \textbf{P1} can be transformed to \eqref{p11}, i.e.,
\begin{align}\label{p11}
\begin{split}
\textbf{P2}:\ &\mathop {\max}\limits_{{\rho_1}^2,\ldots,{\rho_K}^2;{\bf\Theta}_1,\ldots,{\bf\Theta}_K;r_1\ldots,r_K}\sum_{k=1}^{K}r_k\log_2\left( 1+\text{SNR}_k^{SCMU}\right)\\ 
\bf{s.t.}\ \
&\text{C1 - C6}.
\end{split}
\end{align}

Since problem \textbf{P2} is a non-convex optimization problem, it makes the solution difficult to achieve. The alternating-based method is used to solve it, problem \textbf{P2} can be split into two sub-problems as follows.
\subsubsection{Fixed Value of $r_1,\ldots,r_K$} problem \textbf{P2} can be simplified as sub-problem \textbf{P2.1}, where the phase-shift of each RIS element can be optimized to maximize the SNR of each user, which is shown as
\begin{align}\label{p2.1}
\begin{split}
\textbf{P2.1}:\ &\mathop {\max}\limits_{{\rho_k}^2;{\bf\Theta}_k} \text{SNR}_k^{SCMU}\\ 
\bf{s.t.}\ \
&\text{C1 - C3}.
\end{split}
\end{align}
According to \eqref{channel}-\eqref{SNRE}, sub-problem \textbf{P2.1} can be re-written as bellow
\begin{align}\label{p2.1.1}
\begin{split}
\textbf{P2.1.1}:\ &\mathop {\max}\limits_{{\rho_k}^2;\theta_k^1,\ldots,\theta_k^N} \left|\left(h_k+{\bf H}_k {\bf\Theta}_k {\bf G}_k\right)\rho_k\right|^2 \\ 
\bf{s.t.}\ \
&\text{C1 - C3}.
\end{split}
\end{align}
Refer to the existed algorithm which has been proposed in \cite{wu2018intelligent}, by iteratively optimizing one of ${\rho_k}^2$ and $\theta_k^n$ with the other one being fixed at each time, sub-problem \textbf{P2.1} can be simplified and transferred to a solved optimization problem.

For the U$_k$ $\rightarrow$ RIS $\rightarrow$ AP link, the optimal phase-shift of the RIS element $n$ and the optimal transmit power at U$_k$ are given by 
\begin{equation}\label{opt}
\centering
\theta_k^{n^*} = \arg(h_k\rho_k^*)-\arg(H_{k}^n)-\arg(G_{k}^n\rho_k^*),
\end{equation}
and
\begin{equation}\label{opt1}
\centering
\rho_k^* =\sqrt{P-P_{RIS}}\frac{h_k+{\bf H}_k{\bf\Theta}_k^*{\bf G}_k}{\vert h_k+{\bf H}_k{\bf\Theta}_k^*{\bf G}_k\vert},
\end{equation}
where $\arg(\cdot)$ denotes the component-wise phase of a complex. $G_{k}^n$ is the channel from U$_k$ to the RIS element $n$, $H_{k}^n$ is the channel from the RIS element $n$ to AP. $\bf{\Theta}_k^*=\text{diag}\left(e^{j\theta_k^{1^*}},\ldots, e^{j\theta_k^{n^*}}, \ldots, e^{j\theta_k^{N^*}}\right)$ denotes the the optimal phase-shift of the RIS.

\begin{figure}
	\centering{\includegraphics[width=0.5\textwidth]{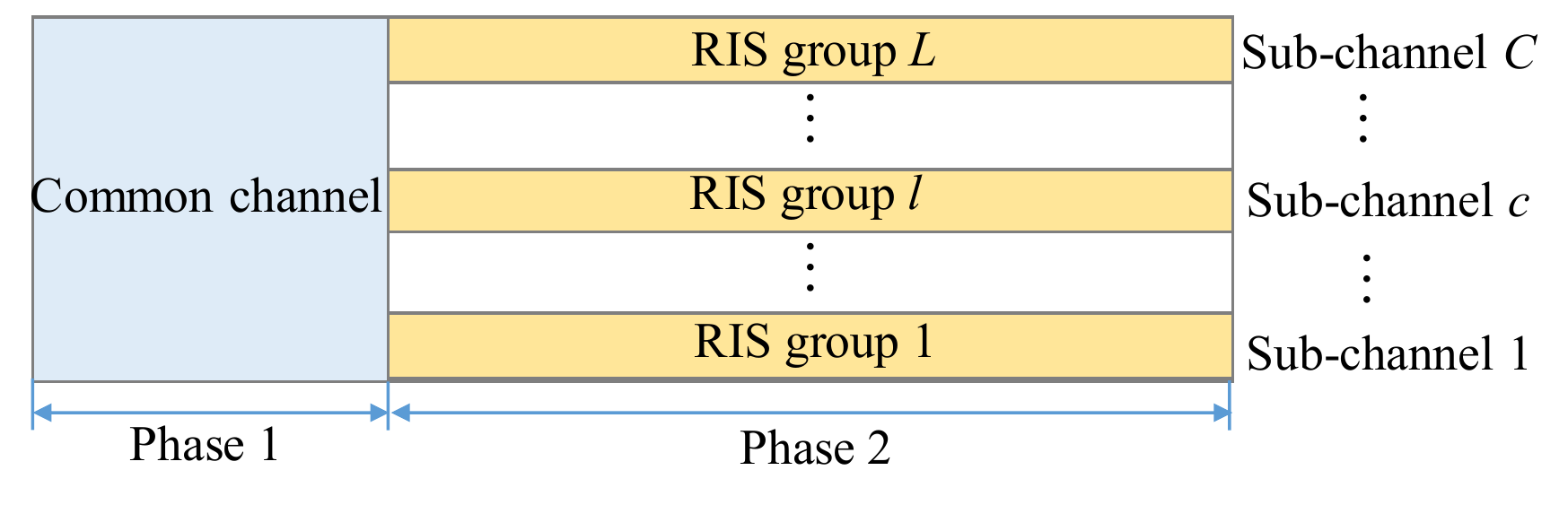}}
	\caption{RIS-assisted MCMU communications, where $L=C>1$. Phase 1 is for negotiation on the common channel, phase 2 is for data transmissions via $L$ RIS groups on $C$ sub-channels that are divided by the common channel.}
	\label{MP}
\end{figure}

\subsubsection{Fixed RIS Configuration ${\bf\Theta}_1,\ldots,{\bf\Theta}_K$ and  ${\rho_1}^2,\ldots,{\rho_K}^2$} problem \textbf{P2} can be rewritten as a sub-problem \textbf{P2.2}.
\begin{align}\label{p2.2}
\begin{split}
\textbf{P2.2}:\ &\mathop {\max}\limits_{r_1,\ldots,r_K} \sum_{k=1}^{K}r_k \\ 
\bf{s.t.}\ \
&\text{C4 - C6}.
\end{split}
\end{align}
Sub-problem \textbf{P2.2} can be viewed as a mixed integer linear program (MILP) problem that can be solved using the branch-and-bound described as in \cite{boyd2004convex}.

\section{RIS-assisted MCMU Communications}\label{sec4}
In this section, we extend RIS-assisted SCMU communications to RIS-assisted MCMU communications, wherein the $C$ sub-channels and the $L$ RIS groups are considered for the multi-user uplink communications system, i.e., $L> 1$ and $C> 1$.

\subsection{Basic Idea}\label{sec4a}
The proposed RIS-assisted MCMU communications are shown in Fig. \ref{MP}, where the RIS elements can be divided into $ L $ groups to support the multiple users' transmissions on the total $ C $ sub-channels. Phase 1 is for negotiation on the common channel, and phase 2 is for data transmissions via $ L $ RIS groups on $ C $ sub-channels. Note that the common channel can be divided into $C$ sub-channels during phase 2, and $L=C>1$. Generally, the adjacent RIS elements are combined into one group by pre-negotiation (i.e., the eRTS/eCTS exchanging) with the AP on the common channel. Each RIS group can be reserved by the different required users to assist the data transmission on the different sub-channels. Different from RIS-assisted SCMU communications, RIS-assisted MCMU communications are presented based on three strategies: the RIS group division, the sub-channel allocation, and the four-dimension reservation. It is noted that the RIS group division is combined with the sub-channel allocation enabling one RIS group to support one user's transmission on one sub-channel. In other words, the reflection coefficients of different RIS groups can be adjusted by the AP for different users to assist multiple data transmissions at a time, thus significantly improving the efficiency of the RIS.

\begin{figure}
	\centering{\includegraphics[width=0.3\textwidth]{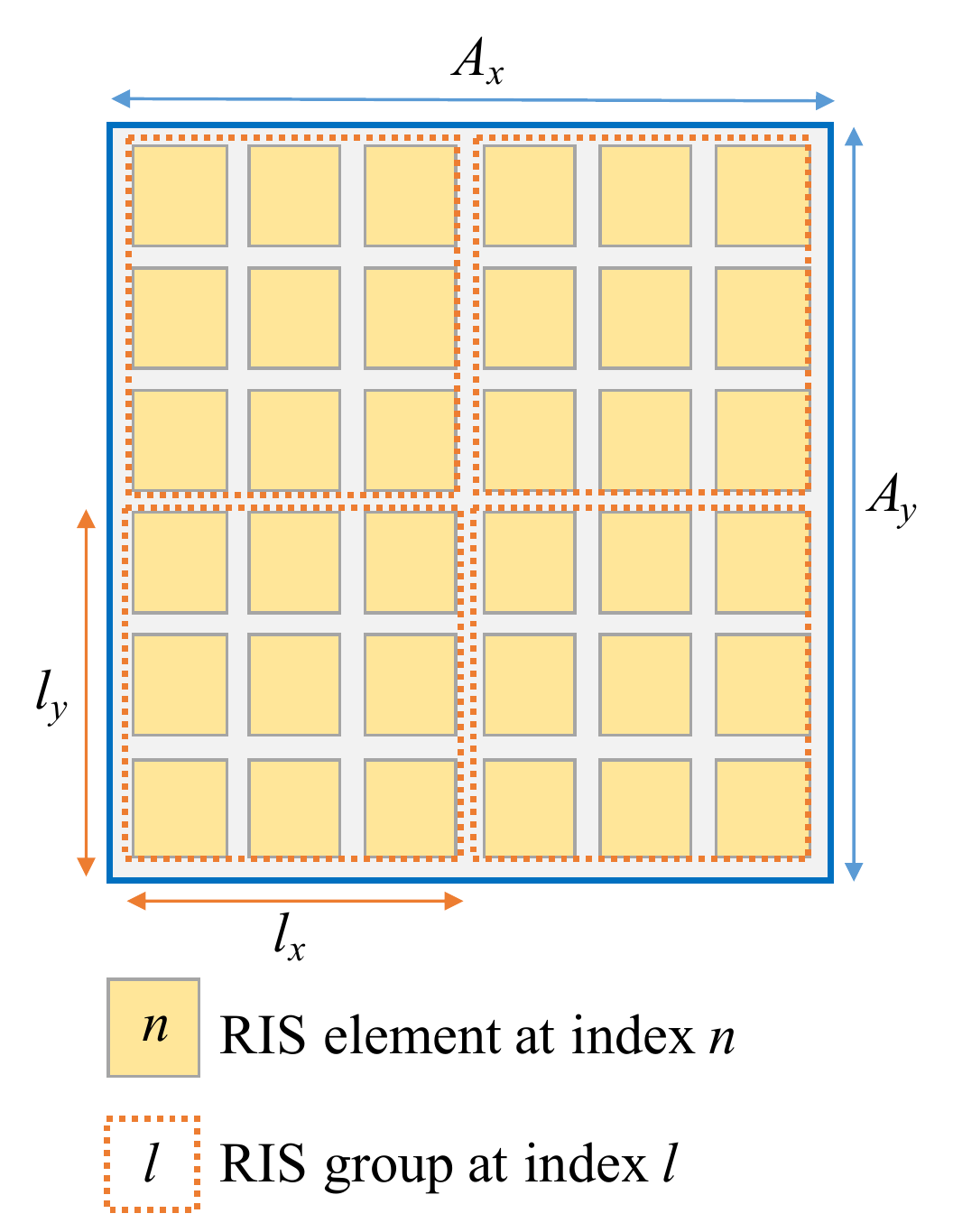}}
	\caption{RIS group division.}
	\label{GRM}
\end{figure}

To implement RIS-assisted MCMU communications, we consider three problems: 1) how to divide the RIS group effectively; 2) how to allocate the common channel and the sub-channel for negotiation and RIS-assisted transmissions; 3) how to achieve the four-dimension reservation.

\subsubsection{RIS Group Division}
In the paper, a group that the adjacent RIS elements can build a small block is illustrated in Fig. \ref{GRM}. Without loss of generality, suppose that the size of the RIS is $A$ ($A_x:A_y$), and the size of each RIS group is $l\ (l_x:l_y)$. Let $L$ denote the number of RIS groups, $1\leq L\leq A$, we have $l=A/L$ (e.g., the elements in the area of $l_x*l_y$ are regarded as one group) and the RIS group ratio is $1/l$. Based on the RIS group division scheme, it is evident that one sub-channel can be occupied by one RIS group. The complexity of RIS elements' reflection coefficients updating for one user in RIS-assisted MCMU communications is $\mathcal{O}(N/L)$. Specifically, if the same reflection coefficient is adopted in one rectangular-shaped group (i.e., each RIS group sets the same reflection coefficient for all the elements included in one RIS group), the AP only needs to adjust the phase-shift of each group but not each element. Therefore, the complexity of RIS phase-shift adjustment can be effectively decreased to $\mathcal{O}(1)$.

\subsubsection{Channel Allocation}
To avoid the collision among the different RIS groups, sub-channels are used for RIS-assisted transmission, where each sub-channel is designated one RIS group (i.e., $L$ = $C$). During the negotiation between user and AP, the eRTS/eCTS transmissions are allowed to execute on the common channel. As the data transmission arrives, the common channel is divided into multiple fixed sub-channels, which enables each RIS group to assist the user's transmission. For example, the available bandwidth is $B$, assuming $C$ sub-channels are required for the multi-user uplink communications system; the available bandwidth of each sub-channel is $B/C$. In other words, each RIS group can occupy the $B/C$ (or $B/L$) bandwidth without interference, and then AP can adjust the phase-shift of each RIS group to assist the different users' data transmissions at different sub-channels.

\subsubsection{Four-Dimension Reservation}
In RIS-assisted MCMU communications, four-dimension reversion is adopted by the requested user and AP. Each dimension reserved zone is denoted by D1 ($r_k$), D2 (${\rho_k}^2$), D3 $(\Theta_k^l)$ and D4 $(c_k,l_k)$.
\begin{itemize}
\item User reservation in phase 1; the requested user reserves the resources of the time zone and the channel zone (e.g., $r_k$ and $(c_k,l_k)$ are for the requested user U$_{k}$). It is noted that the RIS group can be determined once the sub-channel is selected. Besides, the reserved information that is included in the eRTS packet can be directly sent to the AP without the RIS assistance.
\item AP reservation in phase 1; on receiving the reserved request that includes $r_k$ and $(c_k,l_k)$ information from the requested users, the AP makes the decisions as follows: if the time resource and the sub-channel (i.e., the RIS group) are not occupied, the AP computes the optimal transmit power and the optimal reflection coefficient of the reserved RIS group that belongs to the RIS zone. Then, the AP adjusts the phase-shift of the reserved RIS group to the optimal value to assist the requested user's data transmission with the optimal power (e.g., $\Theta_k^l$ is for the requested user U$_k$ relying on the reserved RIS group $l_k$ with the power ${\rho_k}^2$). After this, the reserved information is feedback to the requested user and controller. Otherwise, the AP directly replies to the requested user without the computation of the optimal reflection coefficient.
\end{itemize}

\begin{figure}
	\centering{\includegraphics[width=0.5\textwidth]{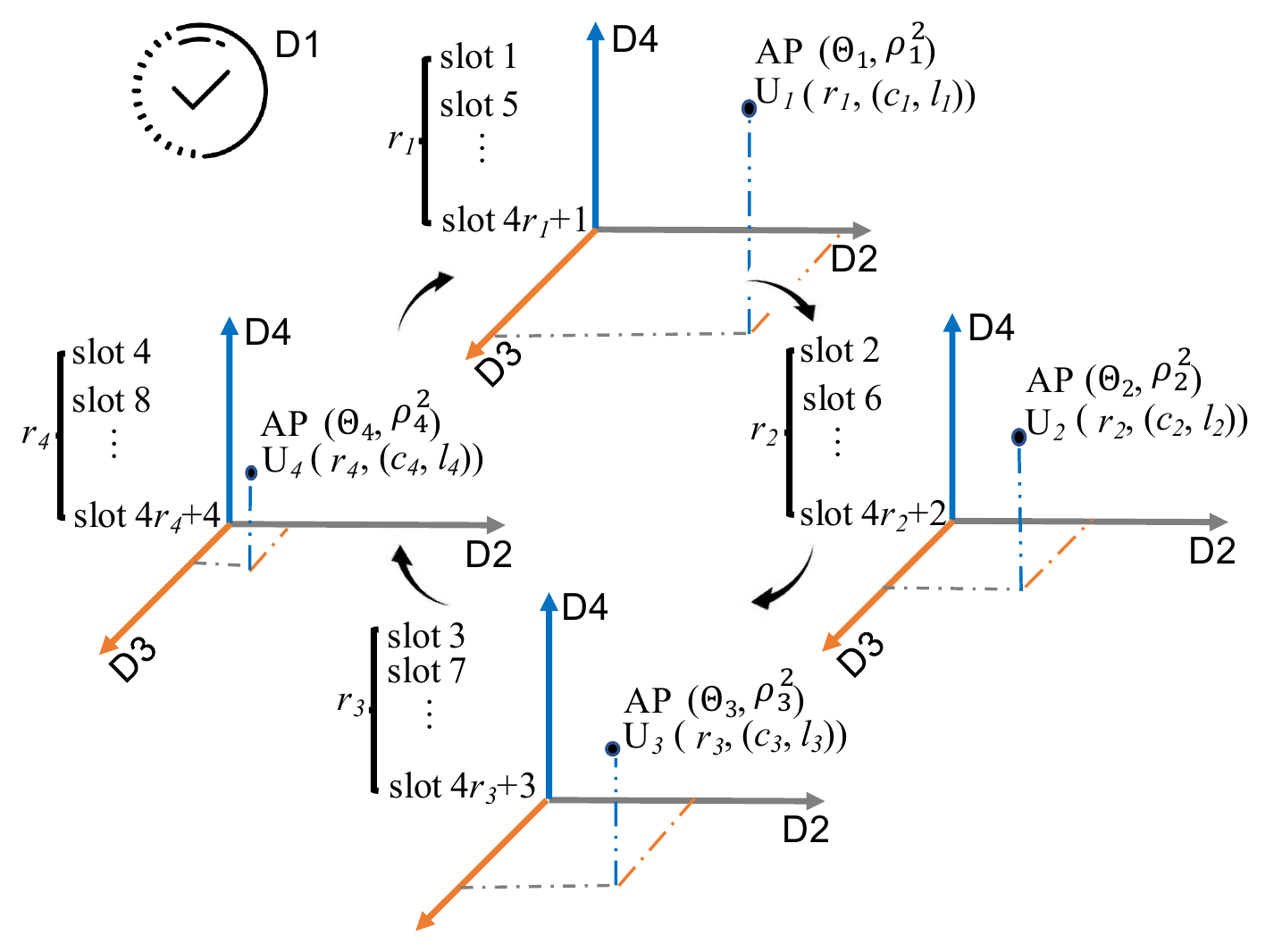}}
	\caption{An illustration of four dimension reversion.}
	\label{FDR}
\end{figure}

An illustration of four dimension revision is displayed in Fig. \ref{FDR}. It is assumed that four users (i.e., U$_1$, U$_2$, U$_3$, and U$_4$) win the negotiation privilege and implement the four-dimension reservation after finishing the backoff, where we set $T_{cycle} = 4$ slots, $\{r_1,r_2,r_3,r_4\}\in [1,r_{max}]$. Since the periodic transmission, the reserved values of three zones comprising D2, D3, and D4 are fixed for one user in each RIS-assisted transmission period, and these values may change in the next frame. For example, in a frame, after negotiation with the AP in phase 1, U$_1$ reserves its slots set, i.e., $\{slot\ 1, slot\ 5, \ldots, slot\ 4r_1+1\}$. In slot 1, the reflect coefficients of elements that are included in the RIS group $l_1$ are adjusted as ${\bf\Theta}_1$, and U$_1$'s transmissions are supported by the RIS group $l_1$ with power ${\rho_1}^2$ on the sub-channel $c_1$. As slot 2 arrives, the reflection coefficients of elements that are included in the RIS group $l_2$ are adjusted as ${\bf\Theta}_2$, and U$_2$'s transmissions are supported by the RIS group $l_2$ with power ${\rho_2}^2$ on the sub-channel $c_2$. Similarly, RIS-assisted transmissions can be successively completed on U$_3$ and U$_4$. When a new periodic start, the transmissions relying on the different RIS group are repeated for U$_1$, U$_2$, U$_3$, and U$_4$, respectively.   

\subsection{Implementation Algorithm}\label{sec4c}
Based on the three strategies mentioned above, the implementation algorithm of RIS-assisted MCMU communications is shown in Alg. \ref{algo1}, where four steps are operated in Phase 1, and three steps are performed in Phase 2.  
\subsubsection{Phase 1: Negotiation Phase}
\begin{itemize}
    \item Step 1: U$_k$ wins the access privilege with backoff operation, and negotiates with the AP on the common channel. 
	\item Step 2: By sending an eRTS packet to the AP at the transmit power $P$, U$_k$ requests ${T_k}^{ini}, r_k, T_{cycle}, c_k$ for its reserved RIS-assisted transmissions. The other users keep silent by setting NAV when U$_k$ is negotiating with the AP. 
	\item Step 3: Once receiving the eRTS packet from U$_k$, the AP judges whether the collision happens on the requested channel and time. If not, the AP estimates the RIS group's phase-shift and calculates transmit power at U$_k$. Then, ${\rho_k}^2$, ${T_k}^{ini}$, $r_k$, $c_k$ and $T_{cycle}$ are recorded into the eCTS packet and is replied to U$_k$. The AP also notifies the RIS controller about $\phi_k^l$, ${T_k}^{ini}$, $r_k$, $c_k$, $l_k$ and $T_{cycle}$. Otherwise, the AP replies U$_k$ without reservation.     
	\item Step 4: Once U$_k$ receiving the eCTS packet from the AP, if RIS group is reserved, U$_k$ ceases its competition and waits for Phase 2. Otherwise, U$_k$ continues its competition and reserves the other RIS group. When the NAV expires, other requested users continue the backoff until Phase 1 terminates.
\end{itemize}

\subsubsection{Phase 2: RIS-assisted Transmission Phase}
\begin{itemize}
     \item Step 1: As Phase 2 arrives, U$_k$ sends the data with the transmit power ${\rho_k}^2$. Meanwhile, the RIS controller sets the phase-shift of the reserved RIS group as $\theta_k^{l(n)}$, and the RIS group assists U$_k$'s data transmission at the reserved sub-channel and the reserved slots.  
	\item Step 2: Once the AP receives the accumulated reflected signal from the RIS group and the LOS's signals from U$_k$, the AP replies an ACK packet to U$_k$.  
	\item Step 3: When the reserved RIS-assisted transmissions between U$_k$ and AP commence, the other users that reserve the same RIS group set NAV to cease their transmissions until the reserved RIS-assisted transmissions of U$_k$ finishes. It is noted that the other users can continue RIS-assisted transmissions on the other RIS group without keeping silent. 
\end{itemize}

\begin{algorithm}[!thb]
    \caption{RIS-assisted MCMU communications}
    \label{algo1}
    \scriptsize
  \KwIn{$C$, $L$\;}
  \KwOut{$\phi_k^l,{\rho_k}^2,c_k,l_k$, ${T_k}^{ini}$ and $r_k$\;}
    Initialize $\vert\phi_k^l\vert = 0$, ${\rho_k}^2=P$, $\forall j, k\in[1,K]$\;
    $t = 0$\;
    \textbf{Phase 1: Negotiation on the common channel}\;
    \If{$t$ is in \textbf{Phase 1}}
    {
        \If{U$_k$ has data to transmit AP }
        {U$_k$ sets its backoff value $\{W, s\}$\;
            \If{$W=0$ }
             {U$_k$ sends eRTS to AP\;
             U$_{j, j\neq k}$ sets NAV and keeps silent\;
              AP receives eRTS packet and judges\; 
                 \If{no collisions} 
                 {AP calculates $\phi_k^l, {\rho_k}^2, l_k$\;
                  AP replies eCTS to U$_k$\;
                  AP notifies the RIS controller about $\phi_k^l$, $l_k$, $r_k$, $c_k$, ${T_k}^{ini}$, $T_{cycle}$\;
                 }
                 \Else
                 {AP replies eCTS without reservation\;
                 }
                U$_k$ receives eCTS packet and waits for \textbf{Phase 2} arriving\;  
             }
              \Else
              {$W--$\; 
              }
      }
         \Else
              {U$_{k}$ keeps silent\;
              }
    }  
         \Else
              {U$_{k}$ waits for \textbf{Phase 1} arriving\;
              }   
     \textbf{Phase 2: RIS-assisted transmission on $L$ groups and $C$ sub-channels}\;   
      \If{$t$ is in \textbf{Phase 2}}
      {U$_{k}$ transmits its data with power, ${\rho_k}^2$, on the reserved sub-channel in the reserved slots\;
       The RIS controller sets $\phi_k^l$ for the RIS group $l_k$\;
       The RIS group $l_k$ adjusts its phase-shift to reflect U$_k$'s signal to AP on the sub-channel $c_k$\;
       AP receives the signals from the RIS group $l_k$ and the LOS path on the sub-channel $c_k$\;
       AP replies an ACK packet to U$_{k}$\; 
      }  
      \Else
              {U$_{k}$ waits for \textbf{Phase 2} arriving\;
              }   
\end{algorithm}

\subsection{Performance Analysis}\label{sec4b}
According to the performance analysis in the sub-section \ref{sec3c}, the proposed Markov model is available for the sub-channel as well. According to (\ref{e2}) and (\ref{eq7}), $\tau_j, j \in [1,C]$ that the stationary probability that one user U$_k$ transmits a data packet in a random slot when $j$ sub-channels are used is provided as in \eqref{e12}. 
\begin{align}\label{e12}
\tau_j=\frac{{2(1 - 2p_j){q}_j}}{{{q}_j[(W + 1)(1 - 2p_j) + p_jW(1 - {{(2p_j)}^m})] + 2(1 - q_j)(1 - p_j)(1 - 2p_j)}}.
\end{align}

Recalling (\ref{e1}), the value of $q_j$ and $p_j$  can be expressed as
\begin{equation}\label{e13}
\centering
q_j = \frac{{{\eta_j}}}{{{\gamma_j} + {\eta_j}}}
\end{equation}
and
\begin{equation}\label{e14}
\centering
p_j = 1 - {(1 - \tau_j )^{K{q_j} - 1}},
\end{equation}
where $q_j$ and $p_j$ denote the probability of a user U$_k$ stays in the ``C” state when $j$ sub-channels are used simultaneously, and the probability that a collision occurs on the common channel when $j$ sub-channels can be used, respectively, $K$ is the number of users in the network.

According to queuing model \cite{bobarshad2009m}, the value of $\gamma_j$ and $\eta_j$ can be obtained as 
\begin{equation}\label{eq17}
\centering
\eta_j  = j\eta
\end{equation}
and
\begin{equation}\label{eq18}
\centering
{\gamma_j} =\frac{\zeta_s^j}{\zeta_e^j\delta+\zeta_s^jt_s +\zeta_c^jt_c},
\end{equation}
where the value of $\eta$ is given in (\ref{rate}).

Therefore, as the single-channel is extended to $j$ sub-channels, the probability of successful transmission ($\zeta_s^j$), the probability of idle channel ($\zeta_e^j$) and the probability of collision ($\zeta_c^j$) can be respectively given as
\begin{equation}\label{Pro}
\centering
\left\{ \begin{array}{l}
\zeta_s^j = Kq_j\tau_j(1-\tau_j)^{Kq_j-1},\\
\zeta_e^j = (1-\tau_j)^{Kq_j},\\
\zeta_c^j = 1-\zeta_s^j-\zeta_e^j.\\
\end{array} \right.
\end{equation}

For RIS-assisted MCMU communications, the process that U$_k$ accesses the sub-channel after contending with other users on the common channel is regarded as an $M/M/C$ queuing model \cite{yang2018channel,cooper2004introduction}, where the U$_k$ successfully contending will be served via one of the $C$ sub-channels. Let $\nu(t)$ denote the number of used sub-channels at time $t$, and the steady-state probability $\pi(j)$ with exactly $j$ number of used sub-channels is defined as
\begin{equation}\label{e20}
{\pi _j} = {\lim _{t \to \infty }}P\{\nu(t) = j\}, \ \ \\ j \ge 0.
\end{equation}
Based on the two-state Markov chain model, it is known that a user accesses the sub-channel with the rate $\gamma_j$ when $j$ sub-channels are used, $j\in[1,C]$. Since ${\sum_{j = 1}^K {\pi_j}}=1$, and then the steady-state probability $\pi_j$ is calculated as
\begin{equation}\label{e21}
{\pi _j} = \frac{{\frac{{\prod\limits_{l = 0}^{j - 1} {{\gamma_l}} }}{{{\eta ^j}j!}}}}{{\sum\limits_{i = 1}^C {\frac{{\prod\limits_{l = 0}^{i - 1} {{\gamma_l}} }}{{{\eta ^i}i!}} + 1} }},   \;\;  j = 1,...,C.
\end{equation}

Given the achievable data rate at each user U$_k$, $R_k^{MCMU}\left(\rho_k,\phi_k \right)=\frac{B}{C}\log_2\left( 1+\text{SNR}_k^{MCMU}\right)$. 
The total capacity of RIS-assisted MCMU communications on all sub-channels can be calculated as follow:
\begin{align}\label{data rrate2}
\begin{split}
\Im_{Total}^{MCMU}&=\sum_{j=1}^{C}\sum_{k=1}^{K}\frac{t_pr_k}{T}\varepsilon_k^j\pi {_j}R_k^{MCMU}\left(\rho_k,\phi_k \right)\\
&=\frac{t_pt_hB}{Tt_sKC}\sum_{j=1}^{C}\zeta_s^j\pi _j\sum_{k=1}^{K}r_k\log_2\left( 1+\text{SNR}_k^{MCMU}\right), 
\end{split}
\end{align}
where $\varepsilon_k^j=\frac{t_h\zeta_s^j}{t_sK}$ is the probability that U$_k$ transmits data in Phase 2 when $j$ sub-channels are used.

\subsection{Performance Optimization}\label{sec4e}
To maximize the system capacity in \eqref{data rrate2}. The performance optimization of RIS-assisted MCMU communications can be separately operated at the AP and users. Specifically, the AP executes the phase-shift optimization of the reserved RIS group for the requested users, and the requested users complete the optimization of sub-channel selection and the number of reserved transmissions. Both of them are formulated in problem \textbf{P3} and problem \textbf{P4}, respectively.  
\subsubsection{The Phase-Shift and The Transmit Power Optimization by AP} problem \textbf{P3} is formulated to solve the optimal phase-shift of each RIS group as
\begin{align}\label{p3}
\begin{split}
\textbf{P3}:\ &\mathop {\max}\limits_{{\rho_1}^2,\ldots,{\rho_K}^2;\phi_1^{l(n)},\ldots,\phi_K^{l(n)}}\sum_{k=1}^{K}\log_2\left( 1+\text{SNR}_k^{MCMU}\right)\\ 
\bf{s.t.}\ \
&\text{C7:}\ \ 0\leq{\rho_k}^2 \leq P-\frac{1}{L}P_{RIS}, \ \forall k,\\
&\text{C8:}\ \ \vert\phi_k^{l(n)}\vert=1,\ n\in l, \ \ \ \forall k, l, \\
&\text{C9:}\ \ \theta_k^{l(n)}\in \mathbf{\Omega},\ n\in l, \ \ \ \ \ \forall k, l,\\
&\text{C10:}\ \ 1\le l \le L, L=C,
\end{split}
\end{align}
where $\text{C7}$ constrains the transmit power of each user when $L$ RIS groups can be used. $\text{C8}$ and $\text{C9}$ separately give the constraints of the amplitude and phase-shift of the RIS group for each user. $\text{C10}$ illustrates the feasibility constraints.

To solve problem \textbf{P3}, we can maximize the SNR of each user by adjusting the phase shift of the selected RIS group. According to \eqref{channel}-\eqref{SNRE}, problem \textbf{P3} can be simplified as problem \textbf{P3.1}, which is shown as
\begin{align}\label{p3.1}
\begin{split}
\textbf{P3.1}:\ &\mathop {\max}\limits_{{\rho_k}^2;\theta_k^{l(n)}} \left|\left(h_k+{\bf H}_k^l {\bf\Theta}_k^l {\bf G}_k^l\right)\rho_k\right|^2 \\ 
\bf{s.t.}\ \
&\text{C7 - C10},
\end{split}
\end{align}
where $G_k^l$ and $H_k^l$ are the channel from U$_k$ to the RIS group $l$, and reflect to AP, $\Theta_k^l$ is the reflection coefficient of the RIS group $l$.

Refer to the existed algorithm which has been proposed in \cite{wu2018intelligent}, by iteratively optimizing one of ${\rho_k}^2$ and $\theta_k^{l(n)}$ with the other one being fixed at each time, the optimal phase-shift of the $n$-th element on the RIS group $l$ and the optimal transmit power at U$_k$ can be given as
\begin{equation}\label{opt2}
\centering
\theta_k^{{l(n)}^*} = \arg(h_k\rho_k^*)-\arg(H_{k}^{l(n)})-\arg(G_{k}^{l(n)}\rho_k^*)
\end{equation}
and
\begin{equation}\label{opt3}
\centering
\rho_k^* =\sqrt{P-\frac{1}{L}P_{RIS}}\frac{h_k+{\bf H}_k^l{\bf\Theta}_k^{l^*}{\bf G}_k^l}{\vert h_k+{\bf H}_k^l{\bf\Theta}_k^{l^*}{\bf G}_k^l\vert},
\end{equation}
where $G_{k}^{l(n)}$ is the channel from U$_k$ to the RIS element $n$ on the group $l$, $H_{k}^{l(n)}$ is the channel from the RIS element $n$ on the group $l$ to AP. ${\bf\Theta}_k^{l^*}=\text{diag}\left(e^{j\theta_k^{1^*}},\ldots, e^{j\theta_k^{l^*}}, \ldots, e^{j\theta_k^{L^*}}\right)$. With considering a same reflect coefficient for all elements in the RIS group $l$, we have $G_{k}^{l(n)}=G_k^l, H_{k}^{l(n)}=H_k^l$, and $\theta_k^{l^*}=\theta_k^{{l(n)}^*}, \forall n\in l$, where $G_k^l$ is the channel from U$_k$ to the RIS group $l$, $H_k^l$ is the channel from the RIS group $l$ to AP. Thus, the optional phase-shift in \eqref{opt2} for the RIS group $l$ can be re-written as
\begin{equation}\label{opt4}
\centering
\theta_k^{{l}^*} = \arg(h_k\rho_k^*)-\arg(H_k^l)-\arg(G_k^l\rho_k^*).
\end{equation}

\subsubsection{The Number of Reserved Transmissions Optimization by User with Sub-Channel Selection} since each user is able to obtain the state of each sub-channel (e.g., the sub-channel is reserved or not) by performing eRTS/eCTS exchanging with AP. Let $V(c_{k})$ represent the sub-channel $c_k$ can be reserved by the U$_k$ or not. In other words, if U$_k$ successfully reserves one sub-channel $c_k$, $V(c_{k})=1$; Otherwise, $V(c_{k})=0$. The problem formulation at each user can be  formulated as 
\begin{align}\label{p4}
\begin{split}
\textbf{P4}:\ &\mathop {\max}\limits_{c_1,\ldots,c_K; r_1\ldots,r_k}\frac{t_pt_hB}{Tt_sKC}\sum_{j=1}^{C}\zeta_s^j\pi _j\sum_{k=1}^{K}r_k\\ 
\bf{s.t.}\ \
&\text{C4, C6},\\
&\text{C11:}\ \ \frac{t_p}{t_s}r_{max}\zeta_s^{C}\leq\frac{t_r}{t_h},\ \\
&\text{C12:}\ \ \sum_{k=1}^{K}V(c_{k}) = t_h\zeta_s^{C}/t_s,\ \forall k,\\
&\text{C13:}\ \ V(c_{k})\in \{0,1\}, \ \forall k\\
&\text{C14:}\ \ 1 \leq c_{k}\leq C, C=L, \ \forall k,
\end{split}
\end{align}
where $\text{C11}$ constrains the RIS-assisted transmission period and the negotiation period when $C$ sub-channels can be used. $\text{C12}$ - $\text{C13}$ constrains the number of sub-channels for users. 

Since $\frac{t_pt_hB}{Tt_sKC}\sum_{j=1}^{C}\zeta_s^j\pi_j$ is a constant when the number of sub-channels is given, problem \textbf{P4} can be rewritten as problem \textbf{P4.1} 
\begin{align}\label{p4.1}
\begin{split}
\textbf{P4.1}:\ &\mathop {\max}\limits_{c_1,\ldots,c_K; r_1,\ldots,r_K} \sum_{k=1}^{K}r_k \\ 
\bf{s.t.}\ \
&\text{C4, C6}, \text{C11-C14}. \\
\end{split}
\end{align}
Sub-problem \textbf{P4.1} can be viewed as a mixed integer linear program (MILP) problem that can be solved using the branch-and-bound described as in \cite{boyd2004convex}.

\begin{figure}[t]
	\centering{\includegraphics[width=0.45\textwidth]{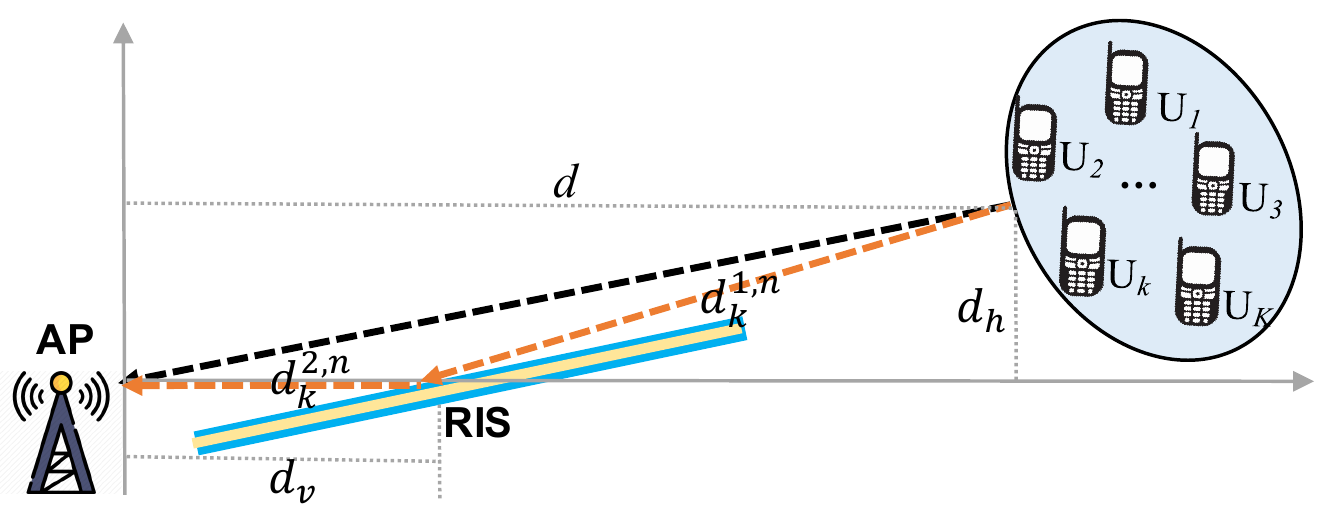}}
	\caption{ The simulated RIS-assisted communications scenario.}
	\label{sim}
\end{figure}

\begin{figure*}[t]
     \centering
    	\subfigure[SNR vs. number of RIS elements]{
    		\begin{minipage}[b]{0.31\textwidth}
    			\includegraphics[width=1\textwidth]{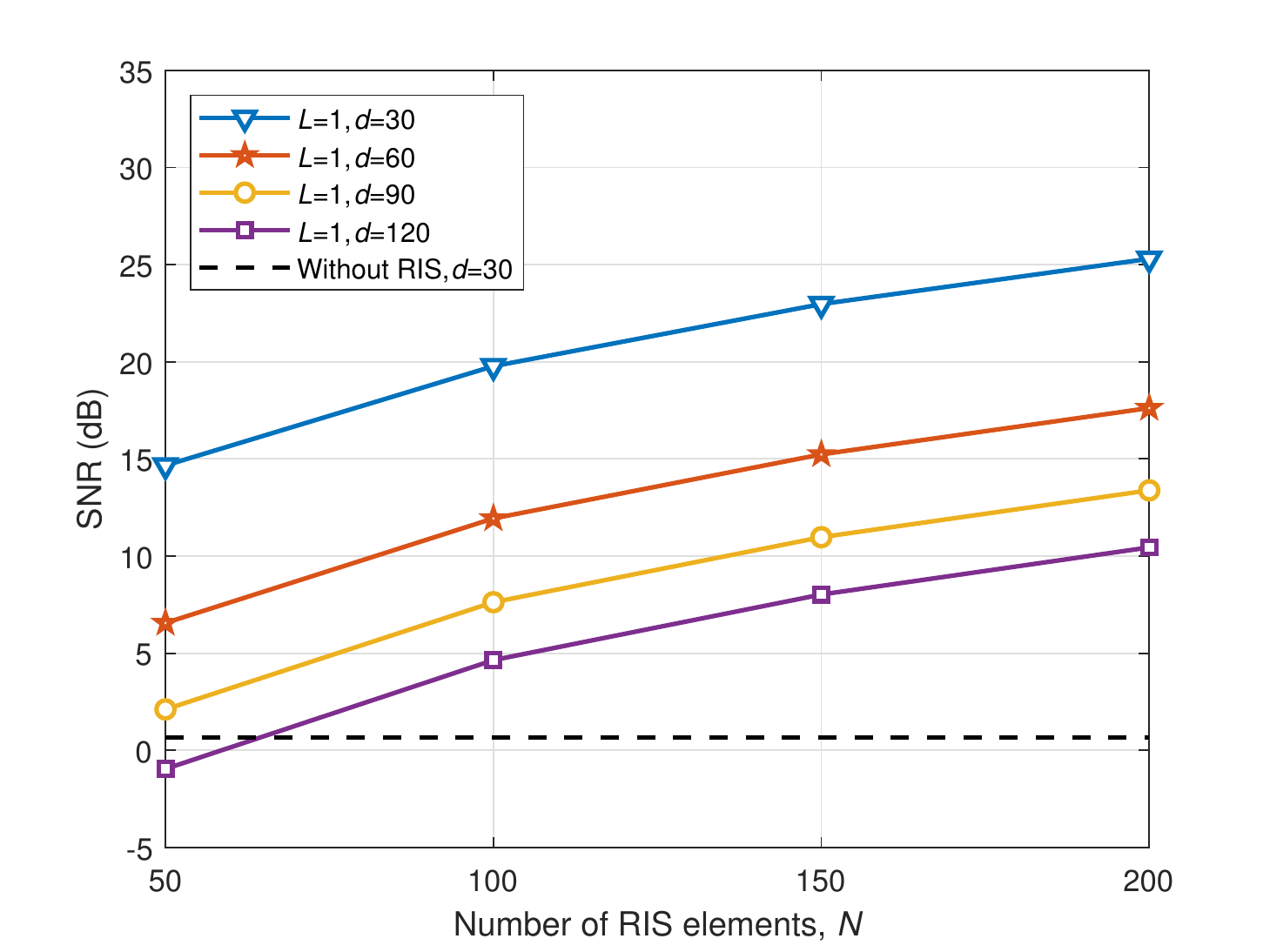}
    		\end{minipage}
    	}
    	\subfigure[Transmit power vs. distance]{
    		\begin{minipage}[b]{0.31\textwidth}
    			\includegraphics[width=1\textwidth]{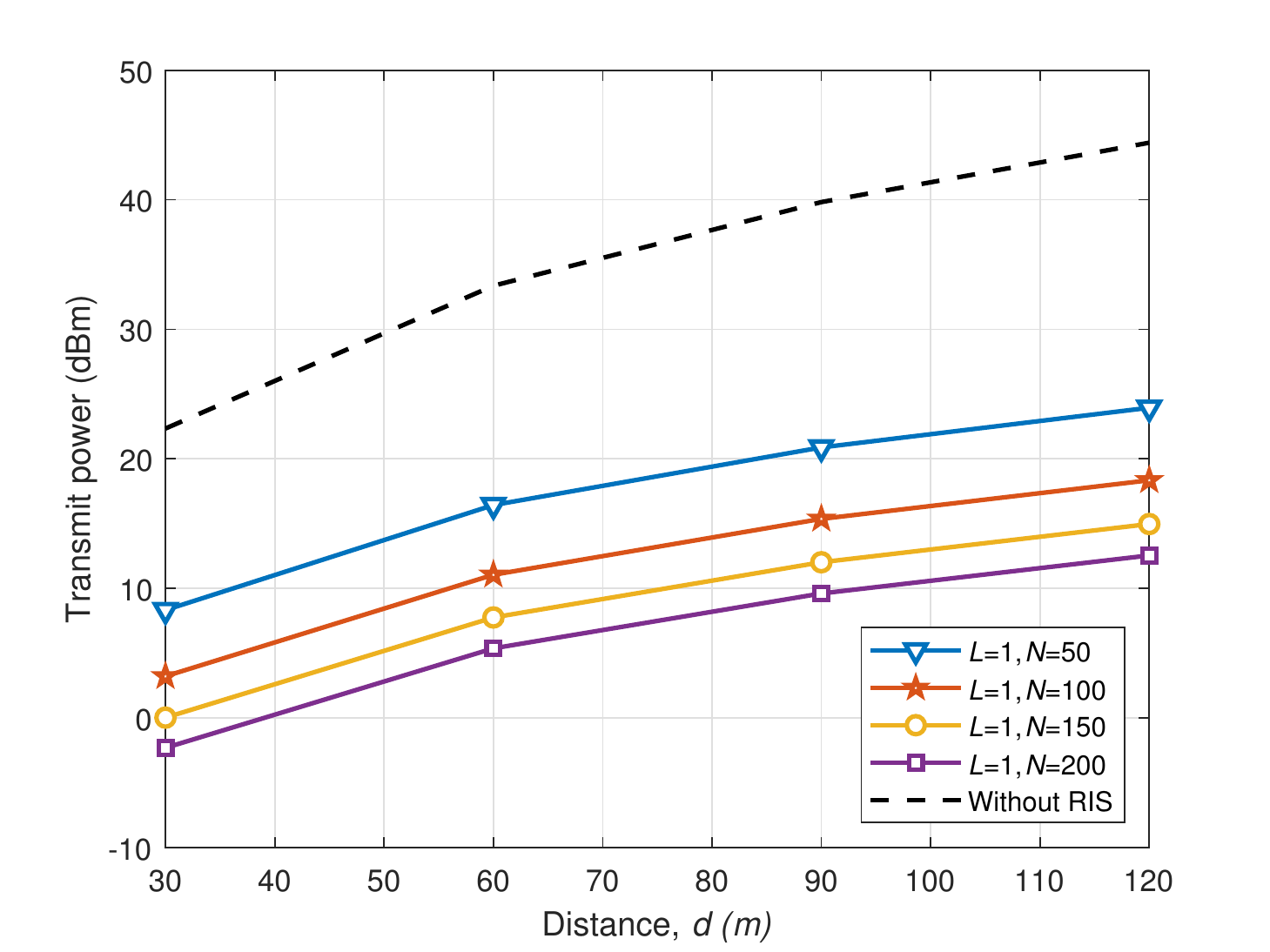}
    		\end{minipage}
    	}
    	\subfigure[Normalized throughput vs. number of users]{
    		\begin{minipage}[b]{0.31\textwidth}
    			\includegraphics[width=1\textwidth]{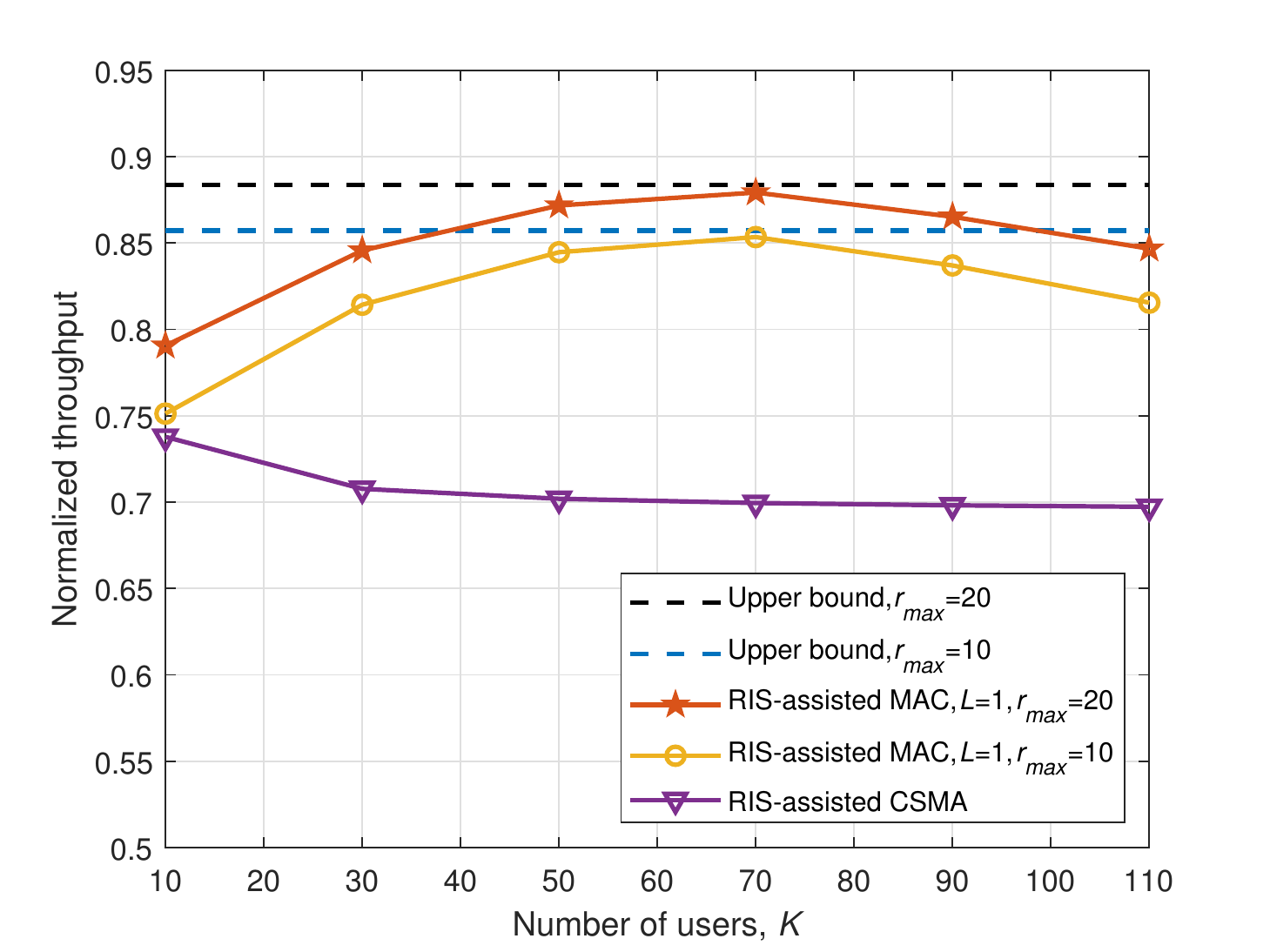}
    		\end{minipage}
    	}
    	\caption{Numerical evaluation of RIS-assisted SCMU communications in terms of SNR, transmit power, and normalized throughput.} 
    	\label{Sim1}   
    \end{figure*}

\section{Performance Evaluation}\label{sec6}
In this section, given a fixed RIS location, we evaluate the performance RIS-assisted MAC with NS-2.35 simulator. Besides, the SimRIS simulator can be used for the RIS evaluation under considering the RIS location \cite{basar2020simris}. RIS-assisted multi-user uplink communications system scenario comprising one AP, one RIS and $100$ users is considered. To simplify the simulation, we assume that all users are with the close position so as to achieve the same SNR at the AP, as shown Fig. \ref{sim}. Wherein $d_k^{1,n}=\sqrt {{d_h}^2+{(d-d_v)}^2}$, $d_k^{2,n}=d_h=2$ m, and $d_v=5$ m. The critical parameters are given as follows: the length of one frame, the time duration of one data transmission, and the transmission cycle are set to be $T=200$ ms, $t_h=20$ ms, $t_p=0.5$ ms, $T_{cycle}=18$ data transmissions, and $r_{max}=20$ respectively. The packet size of eRTS and eCTS are set to be $24$ bytes and $16$ bytes, respectively. SIFS and DIFS are set to be $10$ $\mu$s and $50$ $\mu$s, respectively. The minimum contention window $W_0$ is set to be $15$, and the maximum backoff stage $m$ is set to be $6$. The bandwidth is set to be $B=10$ Mbps, the operating frequency is set to be $f_c=5$ GHz. The number of RIS elements is set to be $N=128$, i.e., $L=1$ and $A(16:8)$. For RIS-assisted MCMU communications, $L$ is valued as ${2, 4, 8, 16}$, where the size of RIS group are set to be $l(8:8), l(8:4), l(4:4), l(4:2)$, respectively. Other required parameters for each user (e.g., U$_k$) are set to be $d=60$ m, $\sigma^2=-80$ dBm, and $P = 5$ dBm (if not specified otherwise).

\subsection {Numerical Results}
In Fig. \ref{Sim1}, the numerical results of RIS-assisted SCMU communications are evaluated in terms of SNR, transmit power, and normalize throughput. Specifically,  Fig. \ref{Sim1} (a) shows the received SNR at the AP versus the number of RIS elements with the different distances between the user and AP. It is observed that the received SNR is increasing with $N$, and it has significant improvement compared with the scheme without RIS. Besides, the received SNR is enhanced when the user is closer to the AP (e.g., the received SNR becomes better as the distance decreases from $120$ m to $30$ m). Fig. \ref{Sim1} (b) shows the transmit power at user versus the distance between the user and AP with the different number of RIS elements. It can be seen that the transmit power at user increases with $d$, and decreases with $N$. Compared to the scheme without RIS, the user's transmit power using the RIS can be decreased significantly, and thus the energy-efficient is achieved in the multi-user uplink communications system. Fig. \ref{Sim1} (c) shows the normalized throughput versus the number of users. As the optimal phase-shift of each RIS element is solved by the AP to support the uplink traffic of each user, RIS-assisted SCMU communications can achieve better throughput performance compared with RIS-assisted CSMA scheme, where RIS-assisted CSMA scheme denotes RIS-assisted transmissions using the conventional CSMA/CA scheme. This is because the negotiation of the RIS resources can avoid interference caused by the RIS reflection. Besides, the throughput performance can be improved as $r_{max}$ is increasing, this is because the increasing of negotiation efficiency. And the upper bounds are given in terms of the different $r_{max}$. 

\begin{figure*}[t]
     \centering
    	\subfigure[SNR vs. number of RIS groups]{
    		\begin{minipage}[b]{0.31\textwidth}
    			\includegraphics[width=1\textwidth]{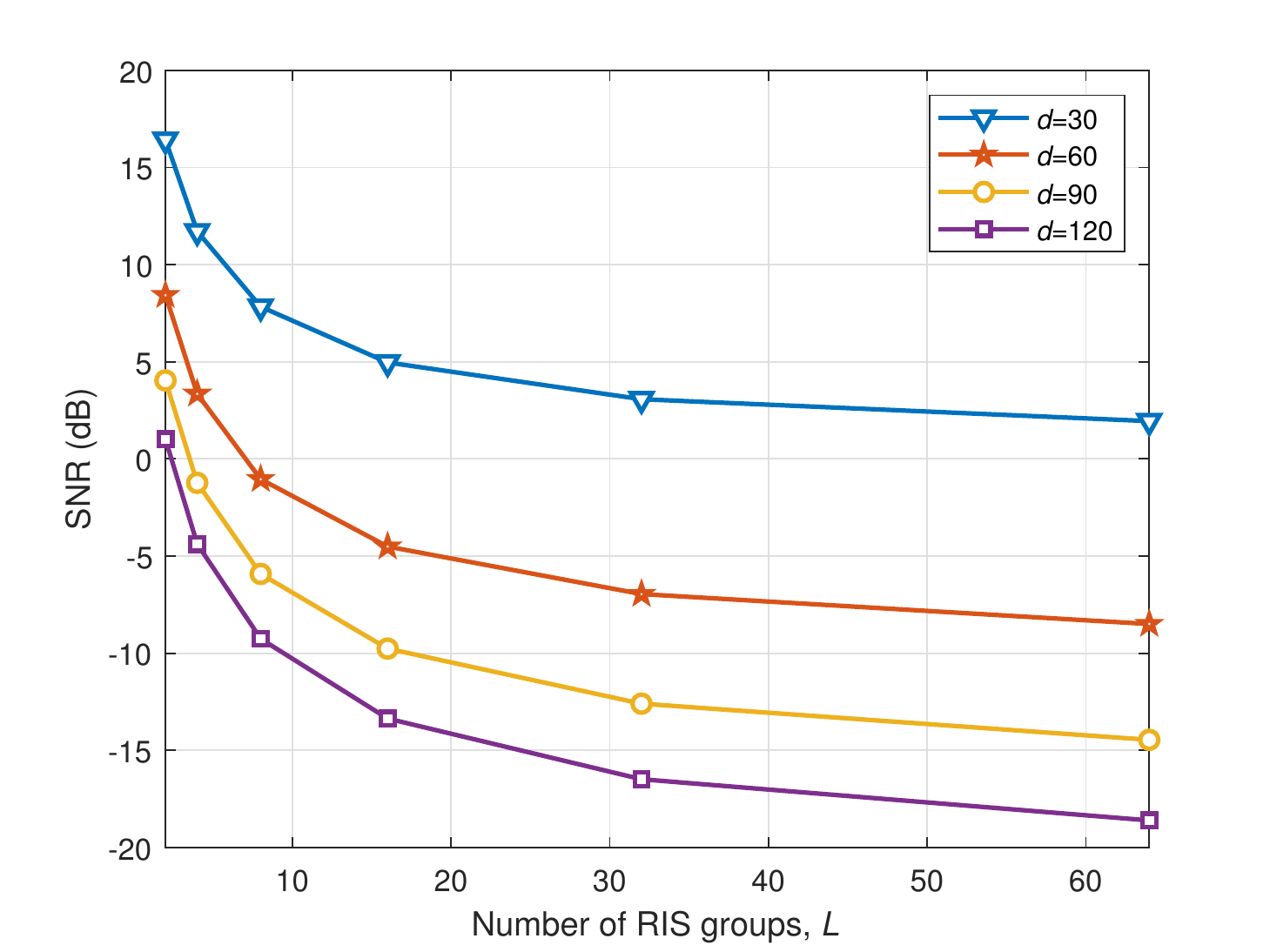}
    		\end{minipage}
    	}
    	\subfigure[Transmit power vs. distance]{
    		\begin{minipage}[b]{0.31\textwidth}
    			\includegraphics[width=1\textwidth]{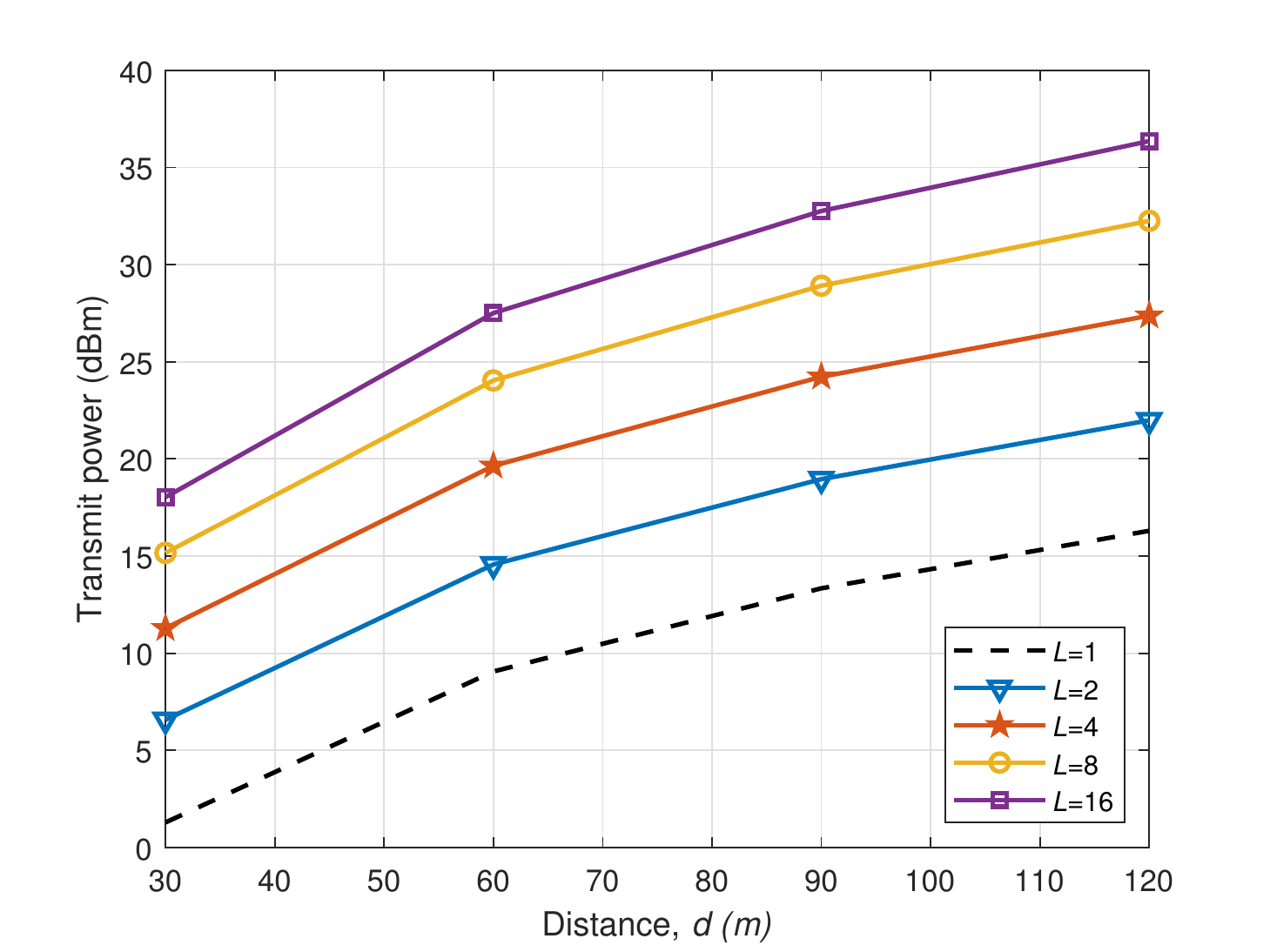}
    		\end{minipage}
    	}
    	\subfigure[Normalized throughput vs. number of users]{
    		\begin{minipage}[b]{0.31\textwidth}
    			\includegraphics[width=1\textwidth]{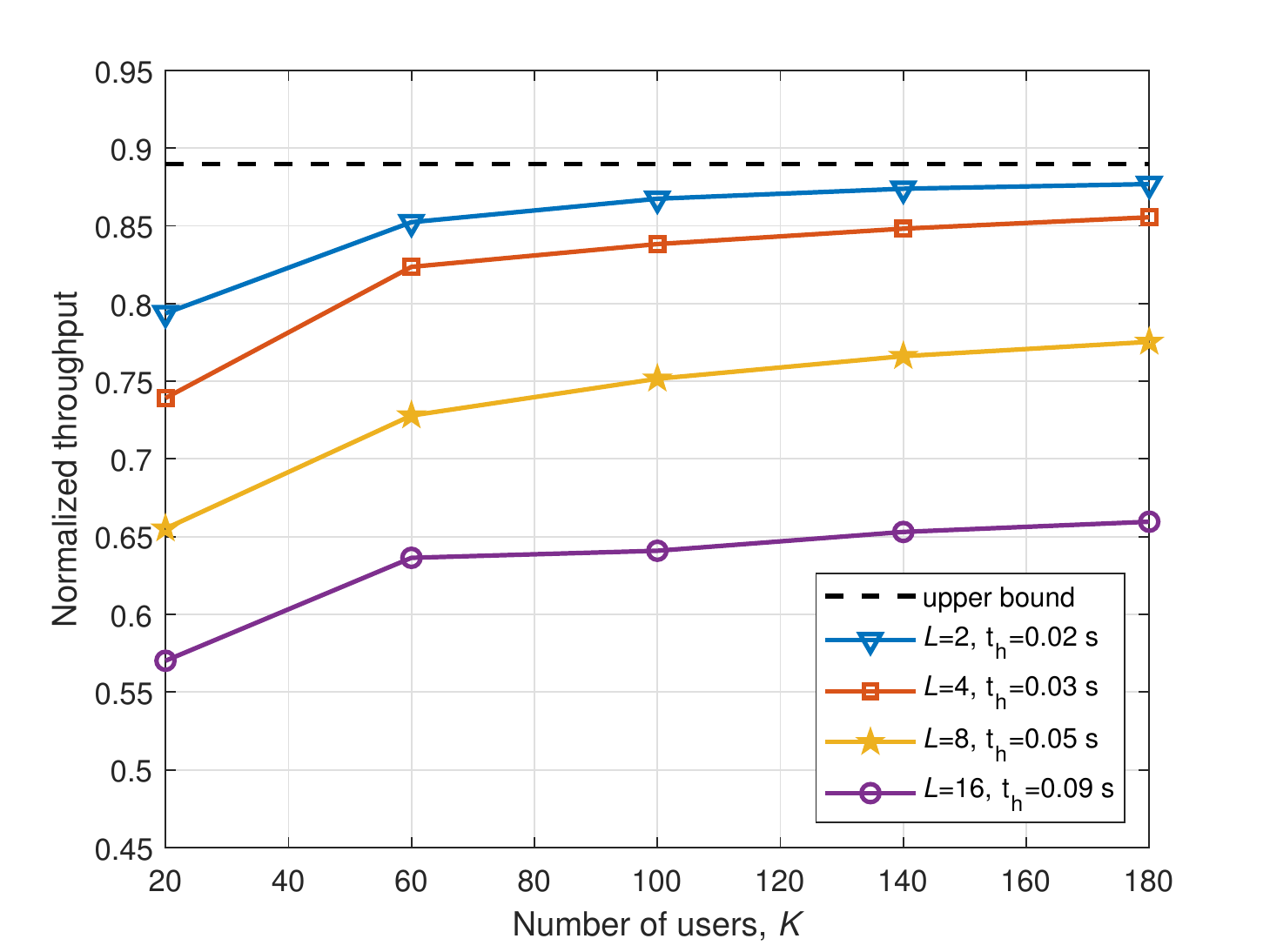}
    		\end{minipage}
    	}
    	\caption{Numerical evaluation of RIS-assisted MCMU communications in terms of SNR, transmit power, and normalized throughput.} 
    	\label{Sim2}
    \end{figure*} 

In Fig. \ref{Sim2}, the numerical results of RIS-assisted MCMU communications are evaluated in terms of SNR, transmit power, and normalize throughput. Specifically, Fig. \ref{Sim2} (a) shows the received SNR at AP versus the number of RIS groups with the different distances between the user and AP. It can be seen that the received SNR decreases with $L$ and $d$, respectively. Compared to RIS-assisted SCMU communications, the SNR of RIS-assisted MCMU communications is not only affected by the distance and but also by the number of RIS groups. Therefore, the SNR of RIS-assisted MCMU communications is lower than RIS-assisted SCMU communications. Fig. \ref{Sim2} (b) shows the transmit power at user versus the distance between the user and AP under the different numbers of RIS groups and sub-channels. It is shown that the transmit power at user increases with $d$ and $L$. Compared with RIS-assisted SCMU communications, the transmit power at the user with RIS group division is improved because that the reflecting elements are divided into the different RIS groups to support the different user's communication at the same time. It indicates that the fewer RIS groups, the less energy consumption. Fig. \ref{Sim2} (c) shows the received SNR at AP versus the number of RIS groups under the different distances between the user and AP. It can be seen that the received SNR decreases with $L$ and $d$, respectively. Besides, the higher SNR can be gained in RIS-assisted SCMU communications compare to RIS-assisted MCMU communications with the RIS group division. As $L$ increases from $2$ to $16$, the normalized throughput decreases from $0.88$ to about $0.65$ after saturation and below the upper bound. It is observed that the value of $t_h$ increases with the $L$ and $K$, this is because the more RIS groups are used to support users' communications, then, the longer duration of the negotiation period is required. Intuitively, the overhead can be quantized as $t_h/(t_h+t_r)=1/(1+t_r/t_h)$. It can be observed that the overhead increases with the ratio of $t_r/t_h$ decreasing. It brings a higher negotiation overhead and leads to a lower transmission efficiency. Therefore, the ratio of $t_r/t_h$ can be dynamically adjusted to control the overhead and the real-time transmission.
\begin{figure*}[t]
     \centering
    	\subfigure[SNR vs. distance]{
    		\begin{minipage}[b]{0.31\textwidth}
    			\includegraphics[width=1\textwidth]{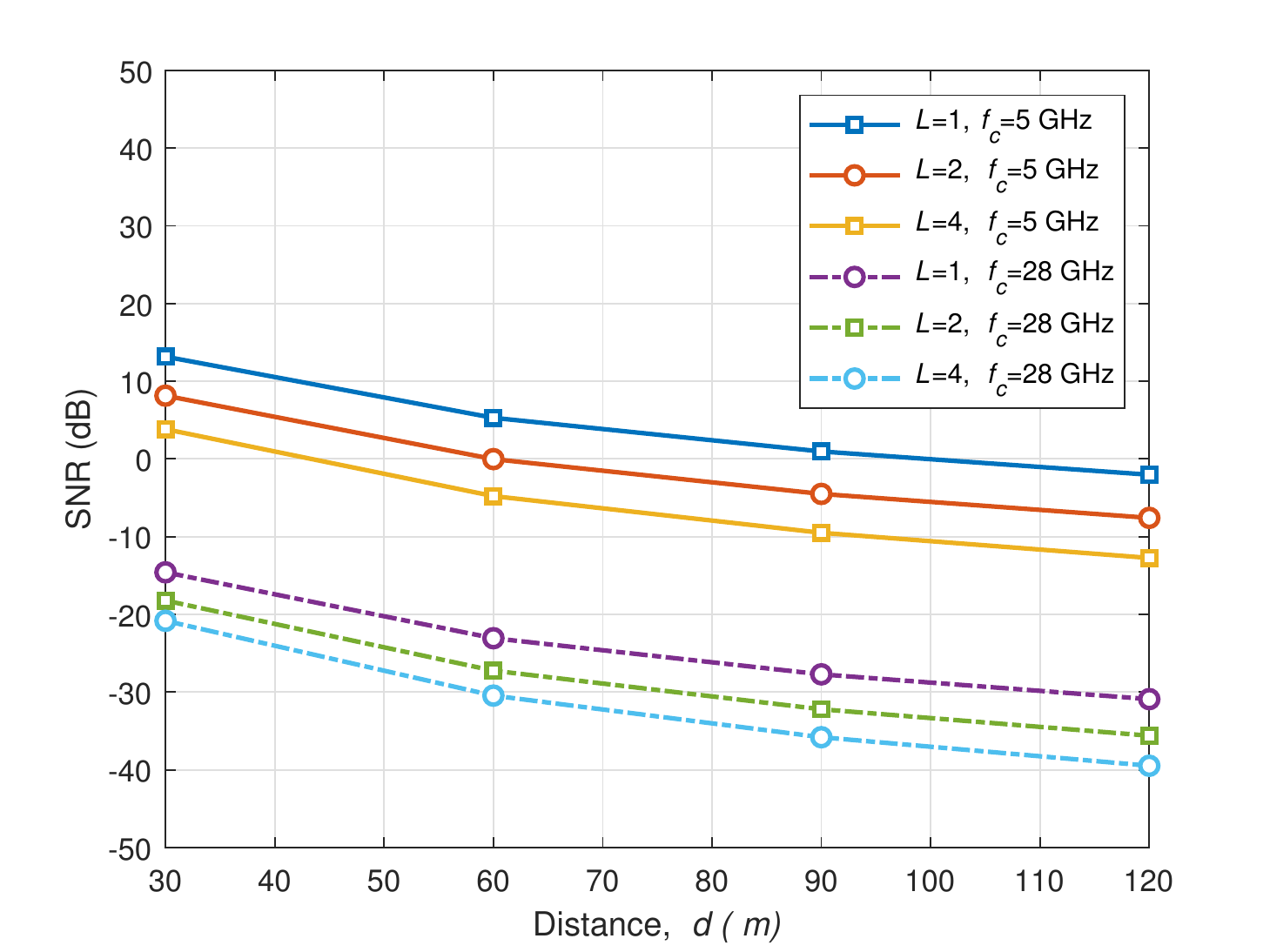}
    		\end{minipage}
    	}   	
    	\subfigure[Total throughput vs. number of users, $f_c=28$ GHz]{
    		\begin{minipage}[b]{0.31\textwidth}
    			\includegraphics[width=1\textwidth]{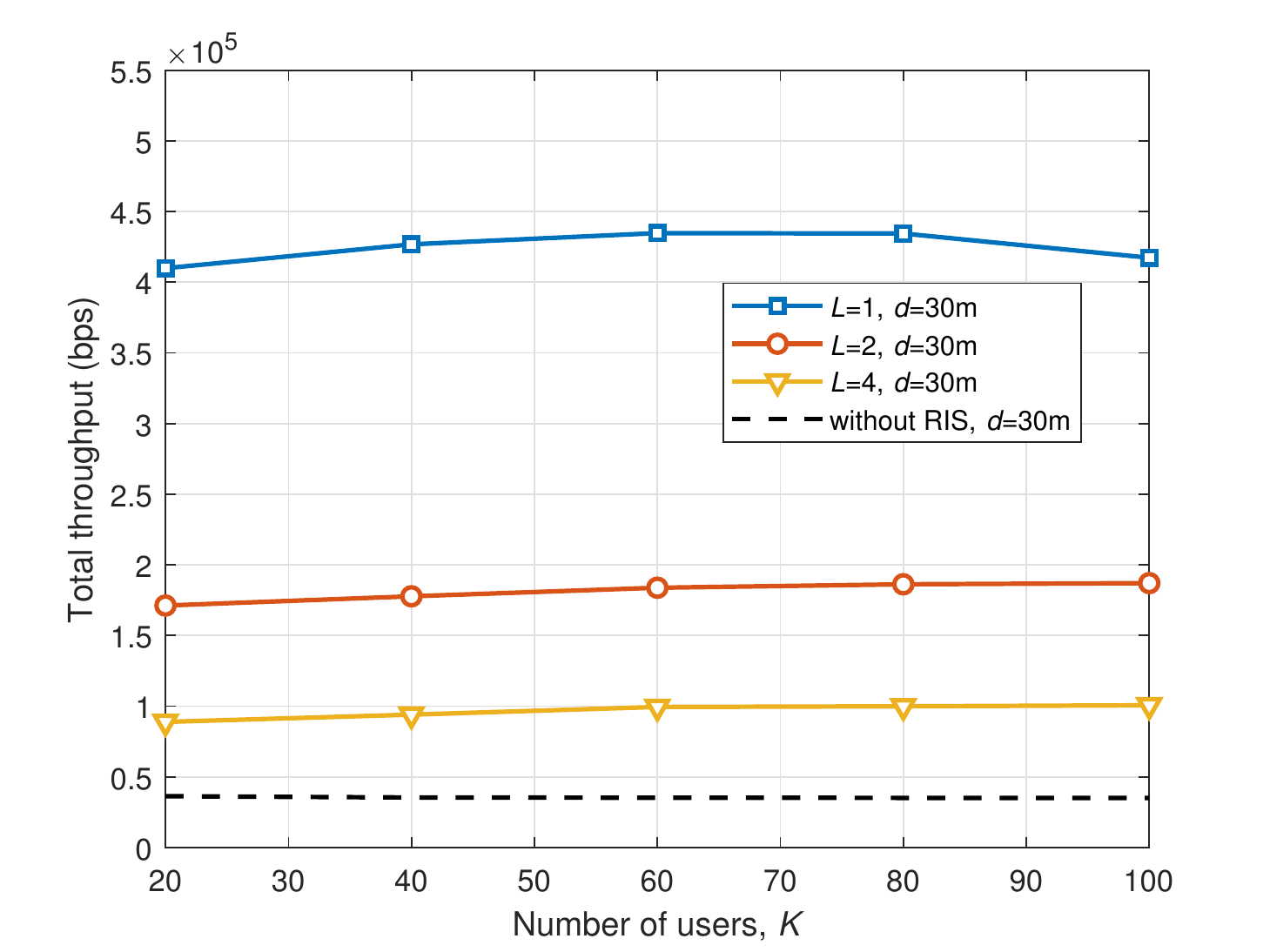}
    		\end{minipage}
    	}
    	\subfigure[Normalized throughput vs. number of users, $f_c=5$ GH]{
    		\begin{minipage}[b]{0.31\textwidth}
    			\includegraphics[width=1\textwidth]{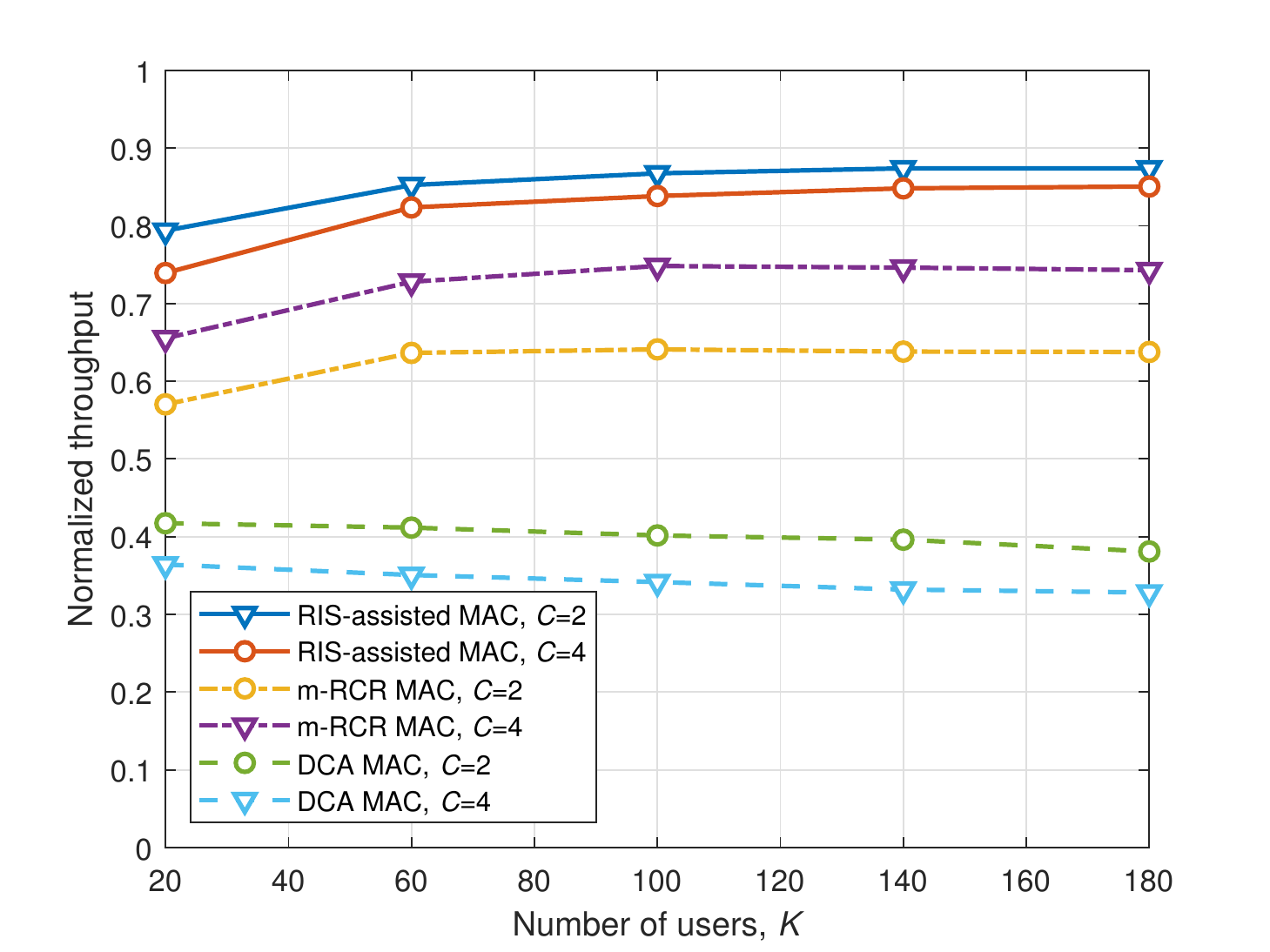}
    		\end{minipage}
    	}
    	\caption{Simulation evaluation of RIS-assisted SCMU/MCMU communications with the different operation frequency, and the various schemes.} 
    	\label{Sim3}
    \end{figure*}  

\subsection {Simulation Results}
In Fig. \ref{Sim3}, simulation evaluations of RIS-assisted SCMU/MCMU communications on the different operating frequencies and the various schemes are given. Fig. \ref{Sim3} (a) evaluates the impact of operating frequency on RIS-assisted communications system, where the operating frequency is set to be $5$ GHz and $28$ GHz, respectively, the number of RIS groups is set to be $1$, $2$ and $4$, respectively. It is shown that the SNR significantly decreases when the operating frequency increases to mmWave frequency due to the severe channel propagation loss. Besides, the similar trends that have been verified in Fig. \ref{Sim2} (a) demonstrate the relationship among SNR, $L$, and $d$ in RIS-assisted SCMU/MCMU communications (i.e., the better received SNR the lower $L$ and $d$). Fig. \ref{Sim3} (b) shows the total throughput in mmWave frequency (i.e., $f_c=28$ GHz), where the total throughput of RIS-assisted mmWave communications first increases and then slightly declines as the number of users increases. Compared with the typical mmWave communications without RIS, the performance of RIS-assisted mmWave communications for each user can be improved up to $50\%$ when the number of RIS groups is set to be $1$. Fig. \ref{Sim3} (c) shows the performance comparison among the different multi-channel MAC protocols in $5$ GHz frequency, where m-RCR MAC protocol \cite{yang2019performance} and DCA MAC protocol \cite{wu2000new} are the benchmarks. Compared with m-RCR MAC and DCA MAC, the proposed RIS-assisted MAC performs the best due to the MDR scheme. In particular, due to the limitation of the negotiation phase in the proposed RIS-assisted MAC, the normalized throughput increases and then keeps almost unchanged as the number of users increases. The performance of m-RCR MAC is better than DCA since the channel reservation is used, and its performance improves with the number of data channels increasing before the control channel saturation. However, its performance keeps almost constant even after the control channel saturation. The DCA MAC performs the worst due to its bottleneck of the control channel, and its channel utilization decreases with the number of data channels after the control channel saturation.

\begin{figure*}[t]
     \centering
    	\subfigure[Collision probability vs. number of users]{
    		\begin{minipage}[b]{0.45\textwidth}
    			\includegraphics[width=1\textwidth]{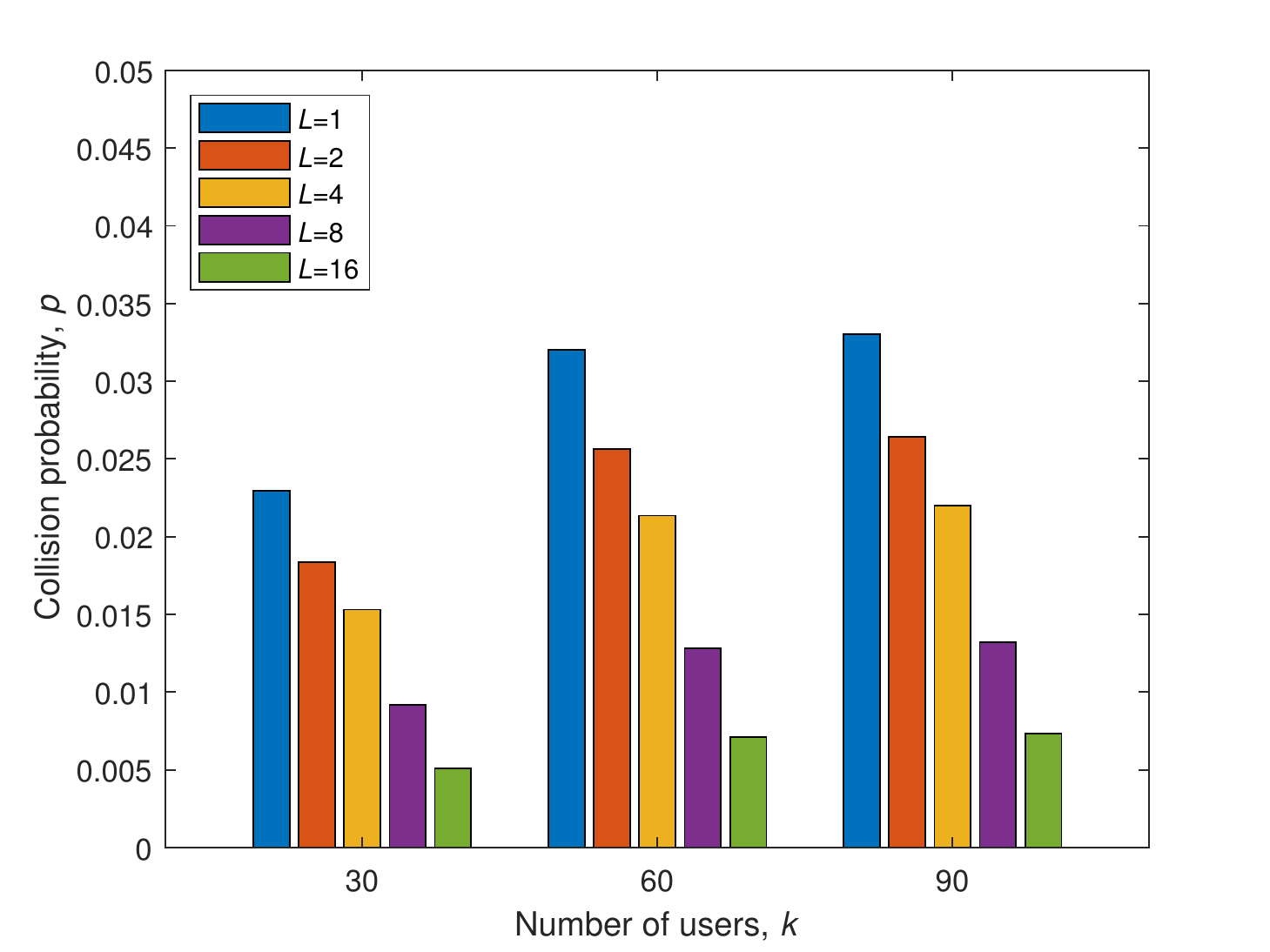}
    		\end{minipage}
    	}
    	\subfigure[Number of served users vs. negotiation period, where $K$=300]{
    		\begin{minipage}[b]{0.45\textwidth}
    			\includegraphics[width=1\textwidth]{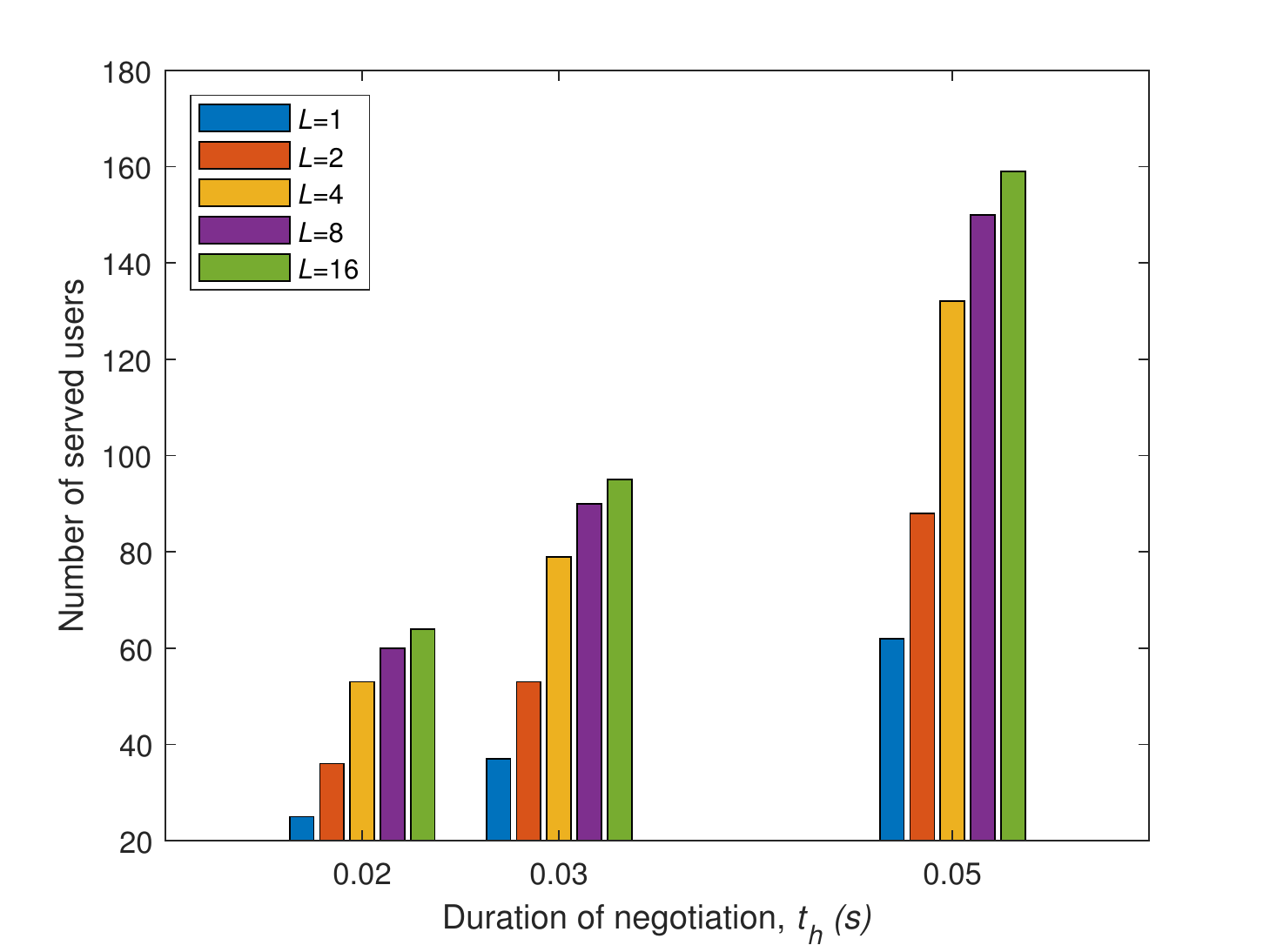}
    		\end{minipage}
    	}
    	\caption{Simulation evaluation of RIS-assisted SCMU/MCMU communications in terms of collision probability and the number of served users.} 
    	\label{Sim4}
    \end{figure*}

In Fig. \ref{Sim4}, simulation evaluations of RIS-assisted SCMU/MCMU communications in terms of the collision probability and the number of served users are presented. Fig. \ref{Sim4} (a) shows the collision probability versus the number of users. It is observed that the collision probability first obviously increases and then slightly climbs as the number of users increases. This is because the collision probability is also affected by the probability that a user is staying at the ``C" state besides the number of users. Moreover, the collision probability also decreases as the number of RIS groups increases due to more users are allowed for accessing at a time. Fig. \ref{Sim4} (b) shows the number of served users versus the duration of the negotiation period in RIS-assisted MCMU communications with a different number of RIS groups, where $K=300$. When the duration of the negotiation period is separately set to be $0.02$ s, $0.03$ s, and $0.05$ s, with the negotiation period enlarging, the more users can be served. Also, the more RIS groups are divided, the more users communicate with AP. Compared with RIS-assisted SCMU communications, RIS-assisted MCMU communications can serve more users, although the system throughput may be decreased. Therefore, there is a tradeoff between the system throughput and the served number of users. 

\vspace{4mm}
\section{Conclusion}\label{sec7}
This paper investigated how to utilize the RIS to assist the multi-user uplink communications system and increase the system's coverage and rate with the power constraint. We proposed an RIS-assisted MAC framework. With this design, RIS-assisted SCMU/MCMU communications were presented, respectively. In particular, RIS-assisted SCMU/MCMU communications can be implemented by two separate phases: the negotiation phase and the RIS-assisted transmission phase. Before RIS-assisted data transmissions, the time slots, the transmit power, the RIS elements, and even the sub-channel can be negotiated in advance. In both cases, RIS elements viewed as the reserved resources can be served for uplink traffics based on the pre-negotiation. It is thereby avoiding interference caused by the RIS reflection and decreasing energy consumption. Besides, RIS-assisted MCMU communications can reduce the RIS computation complexity compared to RIS-assisted SCMU communications. Numerical results on RIS-assisted SCMU/MCMU communications shown that RIS-assisted SCMU communications can gain better throughput, and RIS-assisted MCMU communications can serve more users, and both of them can decrease the transmit power.

In future work, we will address the distributed optimization issues in RIS-assisted networks considering user location. Besides, as the RIS technology is developed in mmWave/THz communications \cite{basar2020simris,CH}, designing RIS-assisted MAC in mmWave/THz frequency is a potential challenge.

\ifCLASSOPTIONcaptionsoff
  \newpage
\fi


\begin{thebibliography}{10}
\providecommand{\url}[1]{#1}
\csname url@samestyle\endcsname
\providecommand{\newblock}{\relax}
\providecommand{\bibinfo}[2]{#2}
\providecommand{\BIBentrySTDinterwordspacing}{\spaceskip=0pt\relax}
\providecommand{\BIBentryALTinterwordstretchfactor}{4}
\providecommand{\BIBentryALTinterwordspacing}{\spaceskip=\fontdimen2\font plus
\BIBentryALTinterwordstretchfactor\fontdimen3\font minus
  \fontdimen4\font\relax}
\providecommand{\BIBforeignlanguage}[2]{{%
\expandafter\ifx\csname l@#1\endcsname\relax
\typeout{** WARNING: IEEEtran.bst: No hyphenation pattern has been}%
\typeout{** loaded for the language `#1'. Using the pattern for}%
\typeout{** the default language instead.}%
\else
\language=\csname l@#1\endcsname
\fi
#2}}
\providecommand{\BIBdecl}{\relax}
\BIBdecl

\bibitem{letaief2019roadmap}
K.~B. Letaief, W.~Chen, Y.~Shi, J.~Zhang, and Y.-J.~A. Zhang, ``The roadmap to
  6G: AI empowered wireless networks,'' \emph{IEEE Communications Magazine},
  vol.~57, no.~8, pp. 84--90, Aug. 2019.

\bibitem{Saad2019vision}
W.~Saad, B.~Mehdi, and M.~Chen, ``A vision of 6G wireless systems:
  Applications, trends, technologies, and open research problems,'' \emph{IEEE
  Network}, vol. 34, no. 3, pp. 134--142, May 2020.

\bibitem{aijaz2020private}
A.~Aijaz, ``Private 5G: The future of industrial wireless,'' \emph{IEEE Industrial Electronics Magazine}, vol. 14, no. 4, pp. 136--145, Dec. 2020.

\bibitem{aijaz2020high}
A.~Aijaz, ``High-performance industrial wireless: Achieving reliable and
  deterministic connectivity over IEEE 802.11 WLANs,'' \emph{IEEE Open Journal
  of the Industrial Electronics Society}, vol.~1, pp. 28--37, Mar. 2020.

\bibitem{huang2020holographic}
C.~Huang, S.~Hu, G.~C. Alexandropoulos, A.~Zappone, C.~Yuen, R.~Zhang,
  M.~Di~Renzo, and M.~Debbah, ``Holographic MIMO surfaces for 6G wireless
  networks: Opportunities, challenges, and trends,'' \emph{IEEE Wireless
  Communications}, vol. 27, no. 5, pp. 118--125, Oct. 2020.

\bibitem{huang2020reconfigurable}
C.~Huang, R.~Mo, and C.~Yuen, ``Reconfigurable intelligent surface
  assisted multiuser MISO systems exploiting deep reinforcement learning,''
  \emph{IEEE Journal on Selected Areas in Communications}, vol. 38, no. 8, pp. 1839--1850, Aug. 2020.  
\bibitem{ntontin2019reconfigurable}
M. Di Renzo et al., ``Reconfigurable intelligent surfaces vs. relaying: differences, similarities, and performance comparison,'' \emph{IEEE Open Journal of the Communications Society}, vol. 1, pp. 798--807, Jun. 2020.

\bibitem{MA}
M. A. ElMossallamy, H. Zhang, L. Song, K. G. Seddik, Z. Han and G. Y. Li, ``Reconfigurable intelligent surfaces for wireless communications: Principles, challenges, and opportunities,'' \emph{IEEE Transactions on Cognitive Communications and Networking}, vol. 6, no. 3, pp. 990--1002, Sep. 2020


\bibitem{di2020smart}
M.~Di~Renzo, A.~Zappone, M.~Debbah, M.-S. Alouini, C.~Yuen, J.~de~Rosny, and S.~Tretyakov, ``Smart radio environments empowered by reconfigurable intelligent surfaces: how it works, state of research, and the road ahead,'' \emph{IEEE Journal on Selected Areas in Communications}, vol. 38, no. 11, pp. 2450--2525, Nov. 2020.

\bibitem{hum2013reconfigurable}
S.~V. Hum and J.~Perruisseau-Carrier, ``Reconfigurable reflectarrays and array
  lenses for dynamic antenna beam control: A review,'' \emph{IEEE Transactions
  on Antennas and Propagation}, vol.~62, no.~1, pp. 183--198, Oct. 2013.

\bibitem{nadeem2019large}
Q. Nadeem, H. Alwazani, A. Kammoun, A. Chaaban, M. Debbah, and M. Alouini, ``Intelligent reflecting surface-assisted multi-user MISO communication: channel estimation and beamforming design,'' \emph{IEEE Open Journal of the Communications Society}, vol. 1, pp. 661--680, May 2020.

\bibitem{di2019smart}
M.~Di~Renzo, M.~Debbah, D.-T. Phan-Huy, A.~Zappone, M.-S. Alouini, C.~Yuen,
  V.~Sciancalepore, G.~C. Alexandropoulos, J.~Hoydis, H.~Gacanin \emph{et~al.},
  ``Smart radio environments empowered by reconfigurable AI meta-surfaces: An
  idea whose time has come,'' \emph{EURASIP Journal on Wireless Communications
  and Networking}, vol. 2019, no.~1, pp. 1--20, Dec. 2019.

\bibitem{di2020hybrid}
B.~Di, H.~Zhang, L.~Song, Y.~Li, Z.~Han, and H.~V. Poor, ``Hybrid beamforming
  for reconfigurable intelligent surface based multi-user communications:
  Achievable rates with limited discrete phase shifts,'' \emph{IEEE Journal on
  Selected Areas in Communications}, vol. 38, no. 8, pp. 1809--1822, Aug. 2020.

\bibitem{mehrotra20193d}
R.~Mehrotra, R.~I. Ansari, A.~Pitilakis, S.~Nie, C.~Liaskos, N.~V. Kantartzis,
  and A.~Pitsillides, ``3D channel modeling and characterization for
  hypersurface empowered indoor environment at 60 GHz millimeter-wave band,''
  in Proc. \emph{IEEE SPECTS}, Berlin, Germany, 2019.

\bibitem{li2019towards}
Z.~Li, Y.~Xie, L.~Shangguan, R.~I. Zelaya, J.~Gummeson, W.~Hu, and K.~Jamieson,
  ``Towards programming the radio environment with large arrays of inexpensive
  antennas,'' in Proc. \emph{NSDI USENIX Symposium}, Boston, USA, 2019, pp. 285--300.

\bibitem{hu2018beyond}
S.~Hu, F.~Rusek, and O.~Edfors, ``Beyond massive MIMO: The potential of data
  transmission with large intelligent surfaces,'' \emph{IEEE Transactions on
  Signal Processing}, vol.~66, no.~10, pp. 2746--2758, Mar. 2018.

\bibitem{basar2019large}
E.~Basar, ``Reconfigurable intelligent surface-based index modulation: A new beyond MIMO paradigm for 6G,'' \emph{IEEE Transactions on Communications}, vol. 68, no. 5, pp. 3187--3196, May 2020.

\bibitem{basar2019transmission}
E.~Basar, ``Transmission through large intelligent surfaces: A new frontier in
  wireless communications,'' in Proc. \emph{IEEE EuCNC}, Valencia, Spain, 2019, pp. 112--117.

\bibitem{zhang2020reconfigurable}
H.~Zhang, B.~Di, L.~Song, and Z.~Han, ``Reconfigurable intelligent surfaces
  assisted communications with limited phase shifts: How many phase shifts are
  enough?'' \emph{IEEE Transactions on Vehicular Technology}, vol.~69, no.~4,
  pp. 4498--4502, Apr. 2020.

\bibitem{di2020practical}
B.~Di, H.~Zhang, L.~Li, L.~Song, Y.~Li, and Z.~Han, ``Practical hybrid
  beamforming with finite-resolution phase shifters for reconfigurable
  intelligent surface based multi-user communications,'' \emph{IEEE
  Transactions on Vehicular Technology}, vol.~69, no.~4, pp. 4565--4570, Feb. 2020.

\bibitem{wu2018intelligent}
Q.~Wu and R.~Zhang, ``Intelligent reflecting surface enhanced wireless network:
  Joint active and passive beamforming design,'' in Proc. \emph{IEEE GLOBECOM}, Abu Dhabi, United Arab Emirates, 2018.


\bibitem{guo2020weighted}
H.~Guo, Y.-C. Liang, J.~Chen, and E.~G. Larsson, ``Weighted sum-rate
  maximization for reconfigurable intelligent surface aided wireless
  networks,'' \emph{IEEE Transactions on Wireless Communications},  vol. 19, no. 5, pp. 3064--3076, May 2020.

\bibitem{huang2019reconfigurable}
C.~Huang, A.~Zappone, G.~C. Alexandropoulos, M.~Debbah, and C.~Yuen,
  ``Reconfigurable intelligent surfaces for energy efficiency in wireless
  communication,'' \emph{IEEE Transactions on Wireless Communications},
  vol.~18, no.~8, pp. 4157--4170, Jun. 2019.

\bibitem{huang2018achievable}
C.~Huang, A.~Zappone, M.~Debbah, and C.~Yuen, ``Achievable rate maximization by
 passive intelligent mirrors,'' in Proc. \emph{IEEE ICASSP}, Calgary, AB, 2018, pp. 3714--3718.

\bibitem{huang2018energy}
C.~Huang, G.~C. Alexandropoulos, A.~Zappone, M.~Debbah, and C.~Yuen, ``Energy
  efficient multi-user MISO communication using low resolution large
  intelligent surfaces,'' in Proc. \emph{IEEE GLOBECOM Workshops}, Abu Dhabi, United Arab Emirates, 2018.

\bibitem{abeywickrama2020intelligent}
S.~Abeywickrama, R.~Zhang, Q.~Wu, and C.~Yuen, ``Intelligent reflecting
  surface: Practical phase shift model and beamforming optimization,''
  \emph{IEEE Transactions on Communications}, vol. 68, no. 9, pp. 5849--5863, Sep. 2020.

\bibitem{han2019large}
Y.~Han, W.~Tang, S.~Jin, C.-K. Wen, and X.~Ma, ``Large intelligent
  surface-assisted wireless communication exploiting statistical CSI,''
  \emph{IEEE Transactions on Vehicular Technology}, vol.~68, no.~8, pp.
  8238--8242, Jun. 2019.

\bibitem{taha2019enabling}
A. Taha, M. Alrabeiah, and A. Alkhateeb, ``Deep learning for large intelligent surfaces in millimeter wave and massive MIMO systems,'' in Proc. \emph{IEEE GLOBECOM}, HI, USA, 2019.

\bibitem{yang2021Int}

B. Yang , X. Cao, C. Huang, C. Yuen, L. Qian, M. Di Renzo, ``Intelligent Spectrum Learning for Wireless Networks with Reconfigurable Intelligent Surfaces,'' \emph{IEEE Transactions on Vehicular Technology}, Feb. 2021.


\bibitem{yu2020robust}
X.~Yu, D.~Xu, Y.~Sun, D.~W.~K. Ng, and R.~Schober, ``Robust and secure wireless
  communications via intelligent reflecting surfaces,'' \emph{IEEE Journal on
  Selected Areas in Communications}, vol. 38, no. 11, pp. 2637--2652, Nov. 2020.

\bibitem{LDai} 
L. Dai,  B. Wang, M. Wang, X. Yang, J. Tan, S. Bi, S. Xu, F. Yang, Z. Chi, M. Di Renzo, C.B. Chae, and L. Hanzo, ``Reconfigurable intelligent surface-based wireless communications: Antenna design, prototyping, and experimental results,'' \emph {IEEE Access}, vol. 8, pp. 45913--45923, Mar. 2020. 
\bibitem{aijaz2013prma}
A.~Aijaz and A.~H. Aghvami, ``A PRMA based MAC protocol for cognitive
  machine-to-machine communications,'' in Proc. \emph{IEEE ICC}, Budapest, 2013, pp. 2753--2758.
\bibitem{shahin2018hybrid}
N.~Shahin, R.~Ali, and Y.-T. Kim, ``Hybrid slotted-CSMA/CA-TDMA for efficient
  massive registration of IOT devices,'' \emph{IEEE Access}, vol.~6, pp.
  18\,366--18\,382, Mar. 2018.

\bibitem{liu2014design}
Y.~Liu, C.~Yuen, X.~Cao, N.~U. Hassan, and J.~Chen, ``Design of a scalable
  hybrid MAC protocol for heterogeneous M2M networks,'' \emph{IEEE Internet of
  Things Journal}, vol.~1, no.~1, pp. 99--111, Feb. 2014.
\bibitem{basar2019wireless}
E.~Basar, M.~Di~Renzo, J.~De~Rosny, M.~Debbah, M.-S. Alouini, and R.~Zhang,
  ``Wireless communications through reconfigurable intelligent surfaces,''
  \emph{IEEE Access}, vol.~7, pp. 116\,753--116\,773, Aug. 2019.

\bibitem{yildirim2019propagation}
I.~Yildirim, A.~Uyrus, E.~Basar, and I.~F. Akyildiz, ``Propagation modeling and
  analysis of reconfigurable intelligent surfaces for indoor and outdoor
  applications in 6G wireless systems,'' \emph{IEEE Transactions on   Communications}, pp. 1--1, Nov. 2020.

\bibitem{you2020channel}
C.~You, B.~Zheng, and R.~Zhang, ``Channel estimation and passive beamforming
  for intelligent reflecting surface: Discrete phase shift and progressive
  refinement,'' \emph{IEEE Journal on Selected Areas in Communications}, vol.~38, no.~11, pp.~2604--2620, 2020.


\bibitem{wei2020parallel}
L.~Wei, C.~Huang, G.~C. Alexandropoulos, and C.~Yuen, ``Parallel factor
  decomposition channel estimation in RIS-assisted multi-user MISO
  communication,'' in Proc. \emph{IEEE SAM Workshop}, Hangzhou, China, 2020.
  
\bibitem{YB}
B. Yang, X. Cao, Z. Han and L. Qian, ``A machine learning enabled MAC framework for heterogeneous internet-of-things networks,'' \emph{IEEE Transactions on Wireless Communications}, vol. 18, no. 7, pp. 3697--3712, Jul. 2019.

\bibitem{cao2018performance}
X.~Cao, B.~Yang, and Z.~Song, ``Performance analysis of hybrid MAC scheme with
  multi-slot reservation,'' \emph{Electronics Letters}, vol.~54, no.~4, pp.
  250--252, Feb. 2018.

\bibitem{bianchi2000performance}
G.~Bianchi, ``Performance analysis of the IEEE 802.11 distributed coordination
  function,'' \emph{IEEE Journal on selected areas in communications}, vol.~18,
  no.~3, pp. 535--547, Mar. 2000.

\bibitem{boyd2004convex}
S.~Boyd, S.~P. Boyd, and L.~Vandenberghe, \emph{Convex optimization}.\hskip 1em
  plus 0.5em minus 0.4em\relax Cambridge university press, 2004.

\bibitem{bobarshad2009m}
H.~Bobarshad and M.~Shikh-Bahaei, ``M/M/1 queuing model for adaptive
  cross-layer error protection in WLANs,'' in Proc. \emph{IEEE WCNC}, Budapest, 2009.

\bibitem{yang2018channel}
B.~Yang, B.~Li, Z.~Yan, and M.~Yang, ``A channel reservation based cooperative
  multi-channel MAC protocol for the next generation WLAN,'' \emph{Wireless
  Networks}, vol.~24, no.~2, pp. 627--646, Feb. 2018.


\bibitem{cooper2004introduction}
R.~Cooper, ``Introduction to queueing theory. 1981,'' \emph{Edward Arnold,
  London}, 2004.

\bibitem{basar2020simris}
E.~Basar, I.~Yildirim, and I.~F. Akyildiz, ``SimRIS channel simulator for
  reconfigurable intelligent surface-empowered communication systems,''
  \emph{arXiv preprint arXiv:2006.00468}, 2020.

\bibitem{yang2019performance}
B.~Yang, B.~Li, Z.~Yan, D.-J. Deng, and M.~Yang, ``Performance analysis of
  multi-channel MAC with single transceiver for the next generation WLAN,''
  \emph{Journal of Network and Computer Applications}, vol. 146, pp. 102408,
 Nov. 2019.

\bibitem{wu2000new}
S.-L. Wu, C.-Y. Lin, Y.-C. Tseng, and J.-L. Sheu, ``A new multi-channel MAC
  protocol with on-demand channel assignment for multi-hop mobile ad hoc
  networks,'' in Proc. \emph{IEEE I-SPAN}, Dallas, USA, 2000, pp. 232--237.

\bibitem{CH}  
C. Huang, Z. Yang, G. C. Alexandropoulos, K. Xiong, L. Wei, Y. Chen, and Z. Zhang,  ``Hybrid beamforming for RIS-empowered multi-hop terahertz communications: A DRL-based method,'' in Proc. \emph {IEEE  Globecom}, Taipei, Taiwan, Dec. 2020.
\end{thebibliography}
\end{document}